\documentclass[twocolumn,10pt,aps,prc,superscriptaddress,nofootinbib]{revtex4-2}
\usepackage{amssymb}
\usepackage{amsmath,bm}
\usepackage{graphicx}
\usepackage[normalem]{ulem}
\usepackage{multirow}
\usepackage[colorlinks,linkcolor=blue,urlcolor=blue,citecolor=blue]{hyperref}
\usepackage{lipsum}
\usepackage[usenames, dvipsnames]{xcolor}
\usepackage{tensor}
\usepackage{isotope}
\usepackage{amsmath}
\usepackage{color}
\usepackage{float}
\allowdisplaybreaks[4]
 
\renewcommand{\sout}{\bgroup \color{red} \ULdepth=-.5ex \ULset}

\usepackage{tikz,xcolor,hyperref}
\definecolor{lime}{HTML}{A6CE39}
\DeclareRobustCommand{\orcidicon}{
	\begin{tikzpicture}
	\draw[lime, fill=lime] (0,0) 
	circle [radius=0.16] 
	node[white] {{\fontfamily{qag}\selectfont \tiny ID}};
	\draw[white, fill=white] (-0.0625,0.095) 
	circle [radius=0.007];
	\end{tikzpicture}
	\hspace{-2mm}
}

\foreach \x in {A, ..., Z}{%
	\expandafter\xdef\csname orcid\x\endcsname{\noexpand\href{https://orcid.org/\csname orcidauthor\x\endcsname}{\noexpand\orcidicon}}
}

\begin{document}

\title{Phase-space excluded-volume approach for light clusters in nuclear medium}

\author{Rui Wang\orcidA{}}
\thanks{Present Address: INFN, Laboratori Nazionali del Sud, I-$95123$ Catania, Italy}
\email{rui.wang@lns.infn.it}
\affiliation{Istituto Nazionale di Fisica Nucleare (INFN), Sezione di Catania, I-$95123$ Catania, Italy}

\author{Zhen Zhang\orcidB{}}
\email{zhangzh275@mail.sysu.edu.cn}
\affiliation{Sino-French Institute of Nuclear Engineering and Technology, Sun Yat-Sen University, Zhuhai 519082, China}

\author{Stefano Burrello\orcidC}
\email[]{burrello@lns.infn.it}
\affiliation{INFN, Laboratori Nazionali del Sud, I-$95123$ Catania, Italy}

\author{Maria Colonna\orcidD}
\email[]{colonna@lns.infn.it}
\affiliation{INFN, Laboratori Nazionali del Sud, I-$95123$ Catania, Italy}

\author{Edoardo G. Lanza\orcidE}
\email{edoardo.lanza@ct.infn.it}
\affiliation{Istituto Nazionale di Fisica Nucleare (INFN), Sezione di Catania, I-$95123$ Catania, Italy}

\date{\today}
\begin{abstract}

A phase-space excluded-volume approach is developed to investigate the in-medium properties of light clusters in nuclear matter.
In this approach, light clusters can exist only if the total nucleon phase-space occupation of the surrounding nuclear medium---including explicit contributions from light clusters---is sufficiently low.
The distribution functions of nucleons and light clusters are determined self-consistently by accounting for the interplay between in-medium effects and thermodynamic properties.
By employing standard Skyrme energy-density functionals to model the nuclear mean-field potential, the approach enables the evaluation of the Mott momentum and the fraction of light clusters in nuclear matter.
Furthermore, it can be readily integrated into dynamical models,
to study in-medium effects on light clusters, based on measured yields in heavy-ion collisions.


\end{abstract}

\maketitle

\newcommand{\red}[1]{\textcolor[rgb]{1.0,0.0,0.0}{\bf \sout{#1}}}
\newcommand{\ora}[1]{\textcolor[rgb]{1.0,0.5,0.0}{\bf #1}}
\newcommand{\gre}[1]{\textcolor[rgb]{0.0,0.6,0.0}{\bf #1}}
\newcommand{\cya}[1]{\textcolor[rgb]{0.0,0.6,1.0}{\bf #1}}
\newcommand{\blu}[1]{\textcolor[rgb]{0.0,0.0,1.0}{\bf (comment: #1)}}

\section{Introduction}
Few-body correlations among nucleons ($N$, including neutrons $n$ and protons $p$) may lead to the formation of light clusters, such as deuterons ($d$) and $\alpha$ particles, in nuclear matter.
Understanding the composition and thermodynamic properties of clustered nuclear matter is of crucial importance for a wide range of topics in both nuclear physics and astrophysics.
These include the structure of finite nuclei~\cite{TypPRC89,YSPRC108}, the dynamics of heavy-ion collisions~\cite{WRPRC110}, the behavior of supernovae explosions~\cite{SumPRC77,ArcPRC78,FisPRC102}, the properties of neutron stars and their mergers~\cite{BurPRC92}, and, more generally, our knowledge of the nuclear equation of state~\cite{SHNPA637,NatPRL104,HemApJ748,TypPRC81,HemPRC91,GulPRC92,OerRMP89,Sum2020}.

Reliable descriptions of these phenomena require careful treatment of the in-medium modifications of light-cluster properties.
In general, the clustering correlations of nucleons embedded in the surrounding nuclear medium can be studied within the thermodynamic Green's function formalism~\cite{RopNPA379,RopNPA399,SchAP202}, where the light-cluster energy is provided by the in-medium many-body Schr\"odinger equation.
Solving this equation yields reduced binding energies of light clusters in the nuclear medium compared with their vacuum values, mainly due to Pauli blocking effects from the surrounding nucleons, including those bound in clusters.
Therefore, the presence of light clusters is suppressed as the density of the medium increases, giving rise to the Mott effect~\cite{MotRMP40}.
This suppression can be effectively incorporated into various mean-field approaches, e.g., by parameterizing the in-medium binding energies and/or by introducing effective meson–cluster couplings.
These methods have been employed to study the composition of nuclear matter~\cite{TypPRC81,VosNPA887,RopPRC92,ZZWPRC95,PaiPRC99,BurEPJA58}, leading to the formulation of several nuclear equations of state with light-cluster ingredients.
We note that studies about nuclear clustering based on {\it ab initio} methods, such as the lattice effective field theory, have also emerged in recent years~\cite{FreRMP90,LBNPRL125,OtsNC13,RZPLB850}.

In addition to theoretical descriptions, significant efforts have also been devoted to extract empirical information on in-medium properties of light clusters from heavy-ion data, such as their Mott densities~\cite{HagPRL108} and abundances~\cite{QLPRL108,BouJPG47,PaiPRL125} in nuclear matter, as well as the effective meson-cluster couplings~\cite{PaiPRL125,CusPRL134}.
Such information, incorporated within mean-field models, is indispensable for the description of heavy-ion collisions and large-scale astrophysical simulations of, e.g., supernova explosion and neutron star merger~\cite{SumPRC77,ArcPRC78,FisPRC102}.

The above empirical studies, based on heavy-ion collision data around the Fermi energy~\cite{BouPRC97}, assume an intermediate-velocity source in global thermal equilibrium, where nucleons and light clusters freeze out.
Since the estimated freeze-out density is very low, 
less than one-third of the nuclear saturation density $\rho_0$, these analyses allowed access to the in-medium properties of light clusters in dilute nuclear matter.
On the other hand, in-medium properties at higher densities may be explored through heavy-ion collisions at higher beam energies.
However, since only local---rather than global~\cite{RamPRL84,ReiPRL92}---thermal equilibrium may be reached in such collisions, 
dynamical approaches, such as transport theories that properly account for in-medium effects on light clusters~\cite{DanNPA533,OnoJPCS420,WRPRC108,CHGPRC109}, are required.

In principle, the in-medium Schr\"odinger equation can be coupled to transport approaches~\cite{BerPR160,AicPR202,BusPR512,WolPPNP125}, such as the Boltzmann-Uehling-Uhlenbeck~(BUU) equation and the quantum-molecular-dynamic approach, to achieve a consistent description of the dynamical evolution of light clusters in nuclear medium.
However, since these dynamical models deal with the time-dependent, spatially-resolved evolution of the nuclear system, the in-medium effects should be treated locally in both space and time.
This involves solving the in-medium Schr\"odinger equation in each spatial cell and at every time step, which is, however, rather computationally demanding.
In order to circumvent this difficulty, a phase-space excluded-volume approach for light clusters has been proposed as an effective way to incorporate the in-medium effects on their dynamical evolution in heavy-ion collisions~\cite{DanNPA533,KuhPRC63,WRPRC108}.
The method captures the main feature of the more sophisticated in-medium Schr\"odinger equation, but with significantly reduced computational cost, by introducing a criterion to determine the survival of light clusters directly from the phase-space density of the surrounding nuclear medium.

In the present work, we generalize the phase-space excluded-volume approach, and employ it to study the properties of clustered nuclear matter.
In particular, we include the contribution of nucleons bound in light clusters in the evaluation of in-medium effects, and consistently account for the interplay between these effects and thermodynamic properties of both nucleons and light clusters--an aspect that is missing in existing models.
The present approach provides a 
viable, effective way for studying in-medium effects on light clusters within mean-field models.
It also serves as a connection  between the in-medium properties of light clusters in nuclear matter---such as their 
abundances---and in-medium effects on their evolution in dynamical processes, thus paving the way for studying the former through light-nuclei observables in intermediate-energy heavy-ion collisions.

We will introduce the phase-space excluded-volume approach for light clusters and the formalism adopted to describe clustered nuclear matter in Sec.~\ref{S:theo}.
The in-medium properties of light clusters, i.e., Mott momenta, Mott densities, and light-cluster fractions in nuclear matter, will be presented in Sec.~\ref{S:rslt}, where their sensitivities to the neutron-proton effective mass splitting will also be explored.
In Sec.~\ref{S:IMSE}, we will compare, for $npd$ matter, the results obtained from the phase-space excluded-volume approach with those by solving the in-medium Schr\"odinger equation.
Conclusions and a brief outlook will be given in Sec.~\ref{S:smy}.
In the Appendix, we outline the equations related to nuclear effective masses within Skyrme energy-density functionals and discuss various possible approximations to the current approach.

\section{Theoretical framework}
\label{S:theo}
\subsection{In-medium Schr\"odinger equation}


For a light cluster of species $\nu$ that consists of $A$ nucleons, with center of mass momentum $\vec{P}$, its wave function $\Psi_{\nu,\vec{P}}(1,2,\cdots,A)$
is the eigenstate of the in-medium Schr\"odinger equation with the corresponding eigenvalue $\epsilon_{\nu}(\vec{P})$, i.e., 
\begin{widetext}
\begin{equation}
\begin{split}
\big[\epsilon_{\tau_1}(\vec{p}_1)+\epsilon_{\tau_2}(\vec{p}_2)&+\cdots+\epsilon_{\tau_A}(\vec{p}_A)- \epsilon_{\nu}(\vec{P})\big]\Psi_{\nu,\vec{P}}(1,2,\cdots,A)\\
& + \sum_{i<j}\big[1-f^{\rm tot}_{\tau_i}(\vec{p}_i)-f^{\rm tot}_{\tau_j}(\vec{p}_j)\big]\int\prod_{l}^{1',2',\cdots, A'}\frac{{\rm d}\vec{p}_l}{(2\pi\hbar)^3}V(ij,i'j')\prod_{k\neq i,j}\delta_{kk'}\Psi_{\nu,\vec{P}}(1',2',\cdots,A') = 0.
\end{split}
\label{E:IMSE}
\end{equation}
\end{widetext}
In the above equation, $i,j,k$ $=$ $1,2,\cdots,A$ label the momentum and isospin state of the constituent nucleons (the spin degrees of freedom can be included analogously), and $\epsilon_{\tau}$ denotes the single-nucleon quasiparticle energies for isospin $\tau$ $=$ $n$ or $p$.
We have simplified the many-body interaction within the $A$-particle cluster to a sum of two-body interactions, i.e., $V(1,2,\cdots,A;1',2',\cdots,A')$ $\approx$ $\sum_{i<j}V(ij,i'j')\prod_{k\neq i,j}\delta_{kk'}$.
Note that $V(ij,i'j')$ vanishes unless $\vec{p}_i+\vec{p}_j$ $=$ $\vec{p}\,'_i+\vec{p}\,'_j$, and one has non-zero $\Psi_{\nu,\vec{P}}(1,2,\cdots,A)$ only when $\vec{p}_1+\vec{p}_2+\cdots+\vec{p}_A$ $=$ $\vec{P}$.
The factor $1-f^{\rm tot}_{\tau_i}(\vec{p}_i)-f^{\rm tot}_{\tau_j}(\vec{p}_j)$ accounts for the Pauli blocking effect of the nuclear medium on the constituent nucleons inside the cluster, with
$f^{\rm tot}_{\tau}(\vec{p})$ representing the total nucleon phase-space occupation of the nuclear medium,
\begin{equation}
f^{\rm tot}_\tau(\vec{p}) = f_\tau(\vec{p}) + f^{\rm clu}_\tau(\vec{p}).
\label{E:ftot}
\end{equation}
In the above, $f_\tau(\vec{p})$ denotes the phase-space occupation of unbound nucleons, and $f^{\rm clu}_\tau(\vec{p})$ denotes that solely from constituent nucleons inside light clusters.
The latter is given by
\begin{eqnarray}
    f^{\rm clu}_n(\vec{p}) & = & \sum_{\nu}^{\rm clu}N\int\frac{{\rm d}\vec{P}}{(2\pi\hbar)^3}f_{\nu}(\vec{P})|\phi_{\nu,\vec{P}}(\vec{p})|^2\label{E:fnclu},\quad\\
    f^{\rm clu}_p(\vec{p}) & = & \sum_{\nu}^{\rm clu}Z\int\frac{{\rm d}\vec{P}}{(2\pi\hbar)^3}f_{\nu}(\vec{P})|\phi_{\nu,\vec{P}}(\vec{p})|^2\label{E:fpclu},
\end{eqnarray}
where $N$ and $Z$ denote the neutron and proton numbers, respectively, of the light cluster species $\nu$, and the summations run over the light-cluster species considered (see also Ref.~\cite{RopPRC79,RopPRC92}).
{In the above equations, $f_\nu(\vec{P})$ represents the occupation of light clusters in momentum space.}
The quantity $|\phi_{\nu,\vec{P}}(\vec{p})|^2$ is the normalized one-body probability distribution of $\Psi_{\nu,\vec{P}}(1,2,\cdots,A)$ in momentum space,
\begin{eqnarray}
   |\phi_{\nu,\vec{P}}(\vec{p})|^2 \propto \int \prod_{i=1}^{A-1}\frac{{\rm d}\vec{p}_{i}}{(2\pi\hbar)^3}|\Psi_{\nu,\vec{P}}(1,2,\cdots,A)|^2,\quad
\end{eqnarray}
where, for brevity, we have omitted the distinction between the neutron and proton components.

The binding energy of the light cluster is defined as the difference between $\epsilon_\nu(\vec{P})$ and the continuum edge of the cluster, i.e., $[N\epsilon_n(\vec{P}/A)+Z\epsilon_p(\vec{P}/A)]$.
A bound-state solution of Eq.~(\ref{E:IMSE}) can only be found when $\epsilon_\nu(\vec{P})$ lies below the continuum edge.
In nuclear matter, this condition defines the Mott momentum $P^{\rm Mott}_\nu$: light clusters of species $\nu$ can exist only when the magnitude of their center-of-mass momentum $|\vec{P}|$ exceeds $P^{\rm Mott}_\nu$.

In perturbative calculations, i.e., for given $f^{\rm tot}_n(\vec{p})$ and $f^{\rm tot}_p(\vec{p})$, Eq.~(\ref{E:IMSE}) can be solved independently.
However, within a fully consistent treatment, Eq.~(\ref{E:IMSE}) is coupled with Eqs.~(\ref{E:ftot})--(\ref{E:fpclu}), and they need to be solved iteratively, which is a significant computational challenge.
Due to this difficulty, for nuclear matter, an effective treatment to consider the light-cluster contribution is usually employed, i.e, to approximate $f_\tau^{\rm tot}(\vec{p})$ in Eq.~(\ref{E:IMSE}) by a Fermi-Dirac distribution normalized to the {\it total} nucleon densities $\rho^{\rm tot}_\tau$.
This uncorrelated-medium assumption decouples the in-medium Schr\"odinger equation from Eqs.~(\ref{E:ftot})--(\ref{E:fpclu}), and reduces the computational complexity substantially.
We evaluate the viability of this assumption in Appendix~\ref{SA:approx}, based on the phase-space excluded-volume approach presented below.

\subsection{Phase-space excluded-volume approach}

Motivated by the copious production of light nuclei in intermediate-energy heavy-ion collisions~\cite{OnoPPNP105},  dynamical descriptions of light clusters within kinetic approaches or transport models have been developed~\cite{DanNPA533,OnoJPCS420,OliPRC99,CocPRC108,WRPRC108,SKJNC15,CHGPRC109}.
In particular, to incorporate in-medium effects on light clusters within kinetic approaches, which is essential for a proper dynamical description, especially considering the transient dense nuclear medium produced in those collisions, a phase-space excluded-volume approach has been proposed~\cite{DanNPA533,KuhPRC63,WRPRC108}.


In this approach, 
a criterion is introduced to
determine the survival of light clusters in the nuclear medium directly from the average nucleon occupation within the phase-space volume corresponding to the free-space wave function of the cluster.
Specifically, a light cluster of species $\nu$ with mass number $A$ and center-of-mass momentum $\vec{P}$ can exist only if both $\langle f_n \rangle_\nu(\vec{P})$ and $\langle f_p \rangle_\nu(\vec{P})$ are less than a cutoff value\footnote{In principle, one can introduce separate cutoff values for $\langle f_n \rangle_\nu(\vec{P})$ and $\langle f_p \rangle_\nu(\vec{P})$.} $F_A^{\rm cut}$, i.e.,
\begin{equation}
\langle f_{\tau}\rangle_{\nu}(\vec{P}) \equiv \int f_{\tau}^{\rm tot}(\vec{p})|\tilde{\phi}_{\nu,\vec{P}}(\vec{p})|^2\frac{{\rm d}\vec{p}}{(2\pi\hbar)^3} < F_A^{\rm cut}.
\label{E:Fcut}
\end{equation}
where $|\tilde{\phi}_{\nu,\vec{P}}(\vec{p})|^2$ denotes the probability distribution $|\phi_{\nu,\vec{P}}(\vec{p})|^2$ in free space, which is usually approximated by a Gaussian form (see Sec.~\ref{S:rslt}).
The phase-space excluded-volume approach essentially describes the Pauli blocking effect exerted by the nuclear medium on the constituent nucleons inside light clusters, which is the main origin of the in-medium effect on light-cluster.
It thus captures the main features of the in-medium Schr\"odinger equation while requiring much less computational effort. 
Most notably, as in the case of the in-medium Schr\"odinger equation, the above criterion defines a Mott momentum.
In nuclear matter, $f_{\tau}^{\rm tot}(\vec{p})$ decreases monotonously as $|\vec{p}|$ increases\footnote{Note that if $P^{\rm Mott}_\nu$ is sufficiently large, $f^{\rm clu}_\tau(\vec{p})$ exhibits a peak around $P^{\rm Mott}_\nu/A$ (see Appendix~\ref{SA:approx}), and it is possible that $f^{\rm tot}_\tau(\vec{p})$ becomes non-monotonic.
We omit this scenario, since it arises only when $F^{\rm cut}_A$ is unreasonably large.}, and consequently, $\langle f_{\tau} \rangle_{\nu}(\vec{P})$ also decreases with increasing $|\vec{P}|$.
Hence, 
the Mott momentum satisfies the following relations
\begin{equation}
    \begin{cases}
    P^{\rm Mott}_{\nu} = 0,~~{\rm if}~\langle f_\tau \rangle_\nu(\vec{P}=0)<F^{\rm cut}_A,~\tau = n,p,\\
    \max\limits_{\tau=n,p}\big[\langle f_\tau\rangle_\nu(P^{\rm Mott}_{\nu})\big] = F^{\rm cut}_A,~~~~~~\rm{if}~P^{\rm Mott}_{\nu} \ne 0. 
    \end{cases}
\label{E:PMt}
\end{equation}

The parameter $F_A^{\rm cut}$ characterizes the strength of the in-medium effect.
A smaller $F_A^{\rm cut}$ 
indicates that the presence of light clusters in the nuclear medium is less favorable, and thus corresponds to a stronger Mott effect.
In principle, different values of $F_A^{\rm cut}$ may be required to better reproduce the realistic Mott momenta and fractions of light clusters for nuclear matter at various densities and temperatures.
Nevertheless, since the density and temperature dependence is mild (as can be seen in Sec.~\ref{S:IMSE}, where, for a simplified case,  
we compare the Mott momentum obtained from the phase-space excluded-volume approach with that from the in-medium Schr\"odinger equation), and one can always focus on a specific density and temperature region in a certain process, we keep $F_A^{\rm cut}$ constant and explore the sensitivity to its choice.

In previous transport-model
studies~\cite{DanNPA533,KuhPRC63,WRPRC108}, the total nucleon occupation $f_\tau^{\rm tot}(\vec{p})$ in Eq.~(\ref{E:Fcut}) is replaced by $f_\tau(\vec{p})$, in the spirit of treating light clusters as completely independent degrees of freedom.
This leads to a smaller $\langle f_{\tau}\rangle_{\nu}(\vec{P})$, and thus the presence of light clusters is less suppressed for a given value of $F_A^{\rm cut}$.
In order to reproduce similar in-medium properties, one may reduce the cutoff parameter to compensate for the omission of the light-cluster contribution to $f^{\rm tot}_\tau(\vec{p})$.
However, from a more consistent point of view, the Pauli blocking originating from nucleons bound in light clusters should be treated on the same footing as that from unbound nucleons, as we aim to do in this work.


Finally, it is important to note that, in the context of heavy-ion collisions, where the phase-space excluded-volume approach was first applied, the total nucleon distribution function used in Eq.~(\ref{E:Fcut}) does not necessarily represent (local) equilibrium conditions. 
In contrast, our aim here is to apply this formalism to equilibrated nuclear matter.
Studying the properties of light clusters in this context---especially in comparison with results from more sophisticated methods---can also help build confidence in extending this approach to out-of-equilibrium scenarios.

\subsection{Nuclear matter with light clusters}

The phase-space distribution functions of nucleons and light clusters in thermal equilibrium are derived resorting to the BUU equation~\cite{UehPR43}.

We first recall how the Fermi-Dirac distribution in uniform nuclear matter can be obtained.
In nuclear matter composed purely of nucleons, 
at thermal equilibrium, the gain and the loss terms in the collision integral of the BUU equation cancel each other, and the nucleon occupation $f_\tau$ satisfies
\begin{equation}
f_{\tau_1}f_{\tau_2}\bar{f}_{\tau_3}\bar{f}_{\tau_4} = f_{\tau_3}f_{\tau_4}\bar{f}_{\tau_1}\bar{f}_{\tau_2},
\end{equation}
where $\bar{f}_\tau$ $=$ $1-f_\tau$ accounts for the Pauli principle.
Dividing both sides by the product of all $\bar{f}_\tau$ and applying the energy conservation condition $\epsilon_{\tau_1}+\epsilon_{\tau_2}$ $=$ $\epsilon_{\tau_3}+\epsilon_{\tau_4}$, we obtain
\begin{equation}
    \frac{f_\tau}{1-f_\tau} \sim e^{-\beta\epsilon_\tau},
\label{E:CE}
\end{equation}
which leads to the Fermi-Dirac distribution for nucleons
\begin{equation}
    f_{\tau}(\vec{p}) = \frac{1}{{\rm exp}\big\{\beta[\epsilon_{\tau}(\vec{p})-\mu_{\tau}]\big\} + 1},
\label{E:FDN}
\end{equation}
with $\mu_{\tau}$ being the nucleon chemical potential and $\beta$ the inverse of the temperature $T$.

For nuclear matter consisting of nucleons and light clusters, since the unbound nucleons can ``feel" the Pauli blocking from the constituent nucleons inside light clusters, 
their occupation deviates from the Fermi-Dirac distribution.
In this sense, an intuitive scenario is to modify $\bar{f}_\tau$ $\rightarrow$ $1-f_\tau-f^{\rm clu}_\tau$ and, correspondingly, Eq.~(\ref{E:CE}), i.e.
\begin{equation}
    \frac{f_{\tau}}{1-f_{\tau}-f^{\rm clu}_{\tau}} \sim e^{-\beta\epsilon_{\tau}},
\label{E:e_clu}
\end{equation}
{We recall that $f_{\tau}^{\rm clu}$ represents the contribution to the  total nucleon occupation $f^{\rm tot}_\tau$ from the nucleons bound in light clusters [see Eqs.~(\ref{E:fnclu}) and (\ref{E:fpclu})]. 
We then obtain the distribution function of unbound nucleons,
\begin{equation}
    f_{\tau}(\vec{p}) = \frac{1-f_{\tau}^{\rm clu}(\vec{p})}{{\rm exp}\big\{\beta[\epsilon_{\tau}(\vec{p})-\mu_{\tau}]\big\} + 1}.
\label{E:fN}
\end{equation}
Adding $f_{\tau}(\vec{p})$ with $f^{\rm clu}_{\tau}(\vec{p})$ leads to the total nucleon occupation
\begin{equation}
\begin{split}
    f^{\rm tot}_{\tau}(\vec{p})  = \frac{1+f^{\rm clu}_{\tau}(\vec{p}){\rm exp}\big\{\beta[\epsilon_{\tau}(\vec{p})-\mu_{\tau}]\big\}}{{\rm exp}\big\{\beta[\epsilon_{\tau}(\vec{p})-\mu_{\tau}]\big\} + 1},
\label{E:ftotNM}
\end{split}
\end{equation}
Under equilibrium conditions, the phase-space distribution function of light clusters $f_{\nu}(\vec{P})$, which is needed in Eqs.~(\ref{E:fnclu}) and (\ref{E:fpclu}) to calculate $f_{\tau}^{\rm clu}$, follows either the Fermi-Dirac or Bose-Einstein distribution, as it can still be derived from the collision integral of the BUU equation, 
when one includes
many-body scatterings involving light clusters~\cite{DanNPA533,KuhPRC63,WRPRC108}, e.g., $Nnp$ $\leftrightarrow$ $Nd$.
The in-medium effects enter the light-cluster distribution function by means of the Mott momentum, namely a light cluster can only exist if the magnitude of its momentum is greater than the Mott momentum.
Finally one has
\begin{equation}
\begin{split}
    f_{\nu}(\vec{P}) = \frac{ H\big(|\vec{P}|-P^{\rm Mott}_{\nu}\big)}{{\rm exp}\big\{\beta[\epsilon_{\nu}(\vec{P})-\mu_{\nu}]\big\}\pm1},
\end{split}
\label{E:fclu}
\end{equation}
where the plus [minus] sign is for fermions [bosons], and $H$ represents the Heaviside step function (equals to 1 for $|\vec{P}|$ $>$ $P^{\rm Mott}_\nu$).

In principle, light-cluster wave functions are needed to evaluate $f^{\rm clu}_{\tau}(\vec{p})$.
Since they are not 
known within the phase-space excluded-volume approach, 
the probability distributions $|\phi_{\nu,\vec{P}}(\vec{p})|^2$ in Eqs.~(\ref{E:fnclu}) and (\ref{E:fpclu}) are approximated by their free-space forms, $|\tilde{\phi}_{\nu,\vec{P}}(\vec{p})|^2$, for simplicity.
Their distinction is found to be small in a variational calculation~\cite{YSPRC108}.
Accordingly, the binding energy (or the mass) of the light clusters, that 
enters the energy
$\epsilon_{\nu}(\vec{P})$, is taken 
as its value in free space.
In principle, one might introduce $\langle f_{\tau} \rangle_{\nu}$-dependent binding energies 
and probability distributions,
to account for their in-medium modifications,
which is a further effect given by the in-medium Schr\"odinger equation.
This will be pursued in the future.

Finally, for given total nucleon density $\rho_{\tau}^{\rm tot}$ and temperature $T$, the nucleon chemical potentials $\mu_n$ and $\mu_p$ can be determined by imposing the relation between $\rho_{\tau}^{\rm tot}$ and the total nucleon
phase-space occupation $f^{\rm tot}_{\tau}(\vec{p})$,
\begin{equation}
g \int \frac{{\rm d}\vec{p}}{(2\pi\hbar)^3} f^{\rm tot}_{\tau}(\vec{p}) = \rho_{\tau}^{\rm tot}
\label{E:NM}
\end{equation}
with $g$ being the nucleon spin degeneracy.
This involves solving the coupled Eqs.~(\ref{E:ftot})--(\ref{E:fpclu}), (\ref{E:PMt}), and (\ref{E:ftotNM})--(\ref{E:NM}) simultaneously, 
with the Mott momentum $P^{\rm Mott}_\nu$ determined self-consistently.
Under chemical equilibrium conditions,
the light-cluster chemical potentials in Eq.~(\ref{E:fclu}) are given by $\mu_\nu$ $=$ $N\mu_n+Z\mu_p$.
Examples of the behavior of
$f_\tau(\vec{p})$, $f^{\rm tot}_\tau(\vec{p})$ and $f^{\rm clu}_\tau(\vec{p})$, obtained in  certain density and temperature conditions, can be found in Appendix~\ref{SA:approx}.

In non-relativistic frameworks, the quasiparticle energy $\epsilon$ can be divided into the kinetic part $\epsilon^{\rm kin}$ and the potential part $\epsilon^{\rm pot}$.
In the present work, the nucleon single particle potential $\epsilon^{\rm pot}_{\tau}(\vec{p})$ is obtained with the standard Skyrme energy-density functional~\cite{ChaNPA627,ChaNPA635} (see Appendix~\ref{SA:Sky}).
The light-cluster quasiparticle potential energies $\epsilon^{\rm pot}_{\nu}(\vec{P})$ are given by scaling $\epsilon^{\rm pot}_{\tau}$ as: 
\begin{equation}
    \epsilon^{\rm pot}_{\nu}(\vec{P}) = N\epsilon^{\rm pot}_n(\vec{P}/A)+Z\epsilon^{\rm pot}_p(\vec{P}/A).
\label{E:scl}
\end{equation}

With the standard Skyrme interaction, in nuclear matter, the momentum dependence of $\epsilon^{\rm pot}_\tau$ can be recast into the kinetic part
leading to 
an effective kinetic energy $\epsilon^{\rm kin,*}_\tau$.
The 
latter is obtained 
by replacing the free-space nucleon mass $m$ in $\epsilon^{\rm kin}_\tau$ with the effective mass $m_{\tau}^*$, defined as
\begin{equation}
    \frac{m}{m^*_\tau} \equiv 1 + m\frac{\vec{p}\cdot\nabla_{\vec{p}}\epsilon^{\rm pot}_\tau(\vec{p})}{p^2}.
\label{E:Em}
\end{equation}
We note that, within standard Skyrme energy-density functionals, the effective mass only depends on the (total) nucleon density and not on the single-particle momentum (see Appendix~\ref{SA:Sky}). 
The remaining (momentum-independent) part of $\epsilon^{\rm pot}_\tau$ is a constant for given $\rho^{\rm tot}_n$ and $\rho^{\rm tot}_p$, and can be absorbed into the chemical potential $\mu_{\tau}$ in Eqs.~(\ref{E:fN}) and (\ref{E:ftotNM}).

The effective mass
of light clusters
can be obtained by applying Eq.~(\ref{E:scl}),
\begin{equation}
\begin{split}
   \frac{m_\nu}{m_\nu^*} &\equiv 1 + m_\nu\frac{\vec{P}\cdot \nabla_{\vec{P}}\epsilon^{\rm pot}_\nu(\vec{P})}{P^2}\\
   &= 1 + \frac{m_\nu}{A^2}\Big[N\frac{\vec{p}\cdot\nabla_{\vec{p}}\epsilon^{\rm pot}_n(\vec{p})}{p^2}+Z\frac{\vec{p}\cdot\nabla_{\vec{p}}\epsilon^{\rm pot}_p(\vec{p})}{p^2}\Big],
\label{E:EmLN}
\end{split}
\end{equation}
with $m_\nu$ being the light-cluster mass in free space and $\vec{p}$ $=$ $\frac{\vec{P}}{A}$.
Note that Eq.~(\ref{E:scl}) also guarantees that the relation $\mu_\nu$ $=$ $N\mu_n$ + $Z\mu_z$ {still} holds after absorbing the momentum-independent part of $\epsilon^{\rm pot}$ into the chemical potentials.

In the present work, we employ the Skyrme interaction MSL$1$~\cite{ZZPLB726} to calculate the quasiparticle potential energies $\epsilon^{\rm pot}_\tau(\vec{p})$ and $\epsilon^{\rm pot}_\nu(\vec{P})$.
In order to explore the influence of the neutron-proton effective mass splitting $m^*_{n-p}$ $\equiv$ $m^*_n-m^*_p$ on light-cluster properties in nuclear matter, we introduce another Skyrme interaction, MSL$1$-m, which 
differs from MSL$1$ in the value of 
$m^*_{n-p}$.
The Skyrme parameters of these two interactions are provided in Appendix~\ref{SA:Sky}.

\section{Results}
\label{S:rslt}

In the present work, we consider the nuclear matter consists of $n$, $p$, $d$, tritons ($t$), helium-3 ($h$) and $\alpha$ particles, within the phase-space excluded-volume approach.
We define three cutoff parameters, namely $F_2^{\rm cut}$ for $d$, $F_3^{\rm cut}$ for $t$ and $h$, and $F_4^{\rm cut}$ for $\alpha$.
To study the sensitivities of light-cluster in-medium properties to the choice of their values, we choose three sets of ${\bf F}^{\rm cut}$ $\equiv$ $(F^{\rm cut}_2,F^{\rm cut}_3,F^{\rm cut}_4)$, as given in Table~\ref{T:Fcut}.
We approximate the free-space one-body probability distribution $|\tilde{\phi}_{\nu,\vec{P}}(\vec{p})|^2$ by a Gaussian form,
\begin{equation}
    |\tilde{\phi}_{\nu,\vec{P}}(\vec{p})|^2 \propto {\rm exp}\Big[-\frac{2\big(\vec{p}-\frac{\vec{P}}{A}\big)^2}{\hbar^2}\sigma^2_{\nu}\Big],
\label{E:Gau}
\end{equation}
with $\sigma_\nu$ determined by the r.m.s. radius of light nuclei in free space, namely, $\sigma_d$, $\sigma_t$, $\sigma_h$ and $\sigma_\alpha$ $=$ $2.26$, $1,59$, $1.76$ and $1.54~\rm fm$, respectively.

\begin{table}[ht!]
\caption{The ${\bf F}^{\rm cut}$ $\equiv$ $(F^{\rm cut}_2,F^{\rm cut}_3,F^{\rm cut}_4)$ used in the present work.}
\begin{tabular}{cccc}
    \hline\hline
    ~~${\bf F}^{\rm cut}$~~ & ~~$F^{\rm cut}_2$~~ & ~~$F^{\rm cut}_3$~~ & ~~$F^{\rm cut}_4$~~ \\
    \hline
    set (i) & $0.10$ & $0.15$ & $0.25$ \\
    set (ii) & $0.20$ & $0.25$ & $0.35$\\
    set (iii) & $0.10$ & $0.15$ & $0.15$\\
    \hline\hline
\end{tabular}
\label{T:Fcut}
\end{table}

\subsection{Mott momenta and Mott densities}
\label{S:PMt}


To illustrate the characteristics of the Mott momenta obtained with the present approach, we consider nuclear matter with different temperatures and isospin asymmetries which are typical in heavy-ion collisions.
Their sensitivity to the strength of the in-medium effect,
i.e., the magnitude of $F_{A}^{\rm cut}$, is also examined.
We present in Fig.~\ref{F:PMt} the $P^{\rm Mott}_\nu$ for $d$, $t$, $h$ and $\alpha$ as a function of the baryon density $\rho_B$ $=$ $\rho^{\rm tot}_n+\rho^{\rm tot}_p$, for symmetric nuclear matter obtained with ${\bf F}^{\rm cut}$ (i) and (ii).
We also include in Fig.~\ref{F:PMt} the results for asymmetric matter with isospin asymmetry $\delta$ $\equiv$ $(\rho^{\rm tot}_n-\rho^{\rm tot}_p)/\rho_B$ $=$ $0.13$, employing ${\bf F}^{\rm cut}$ (i), for comparison.
For all 
cases, two different temperatures $T$ $=$ $5~\rm MeV$ and $20~\rm MeV$ are considered.

\begin{figure}[hb]
\centering
\includegraphics[width=\linewidth]{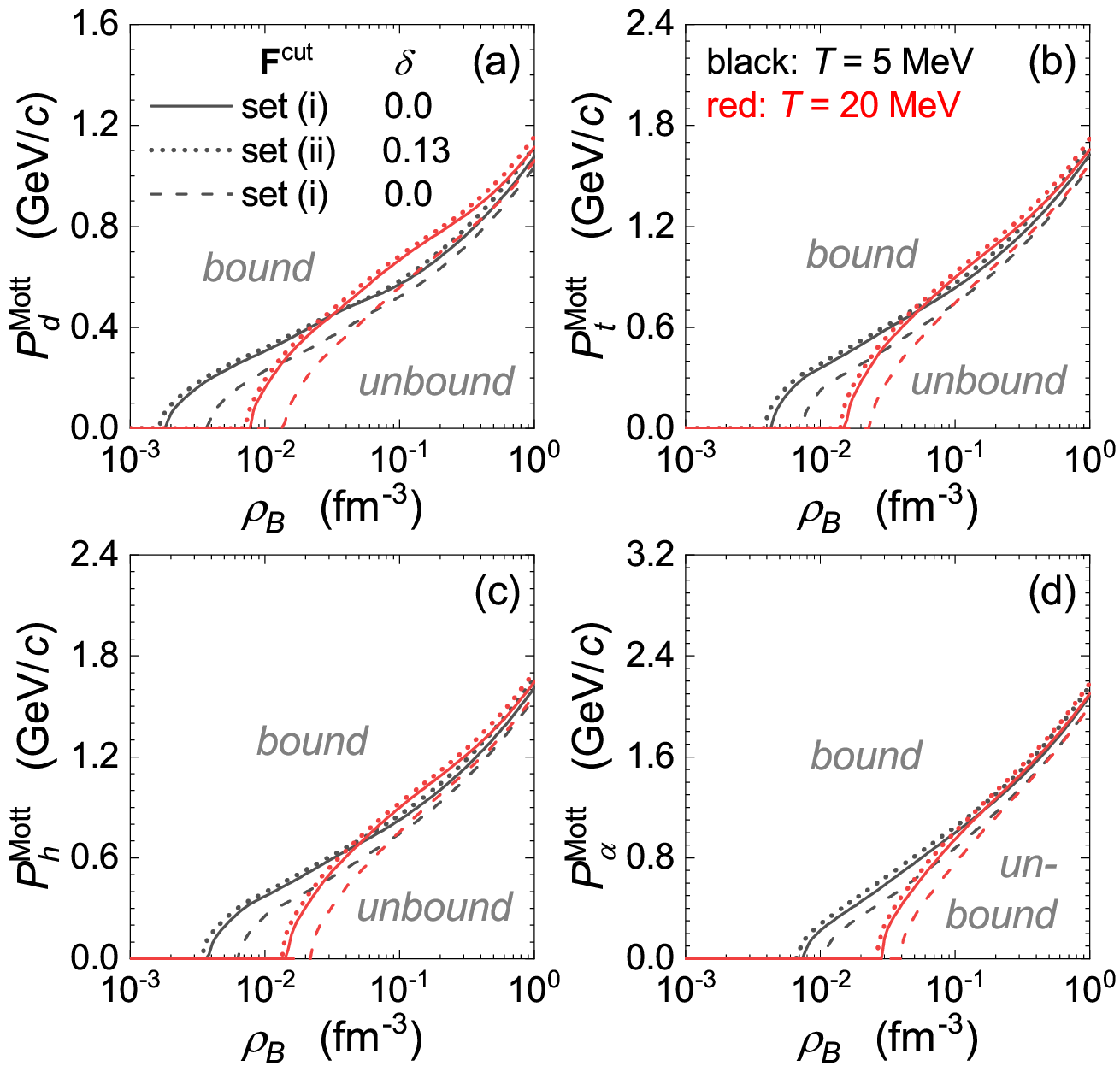}
\caption{Mott momenta $P^{\rm Mott}_\nu$ for (a) $d$, (b) $t$ , (c) $h$ and (d) $\alpha$ as functions of baryon density $\rho_B$ obtained with the phase-space excluded-volume approach for symmetric nuclear matter with ${\bf F}^{\rm cut}$ (i) $=$ $(0.10,0.15,0.25)$ and (ii) $=$ $(0.20,0.25,0.35)$, and for nuclear matter with isospin asymmetry $\delta$ $=$ $0.13$ with ${\bf F}^{\rm cut}$ (i), at temperature $T$ $=$ $5~\rm MeV$ and $20~\rm MeV$, respectively.
All calculations are obtained by employing the Skyrme interaction MSL$1$.
}
\label{F:PMt}
\end{figure}

From Fig.~\ref{F:PMt}, we notice that $P^{\rm Mott}_\nu$ becomes non-zero starting from a certain $\rho_B$.
This defines the Mott densities $\rho^{\rm Mott}_\nu$ of light clusters, above which light clusters can reside only in part of the whole momentum space.
The overall larger ${\bf F}^{\rm cut}$ (ii), which corresponds to a weaker in-medium effect, allows light clusters with a higher $\langle f_\tau \rangle_{\nu}(\vec{P})$ to bind, thereby decreasing the Mott momenta.
A positive $\delta$ leads to a larger neutron density and subsequently a larger $\langle f_n \rangle_{\nu}(\vec{P})$. 
It thus increases the Mott momenta, though the effect is not that pronounced.
As the temperature increases, the total nucleon occupation $f^{\rm tot}_\tau(\vec{p})$ exhibits a broader momentum distribution.
Therefore, at $T$ $=$ $20~\rm MeV$, we obtain smaller $P^{\rm Mott}_\nu$ for lower densities and larger $P^{\rm Mott}_\nu$ for higher densities, with respect to the results at $T$ $=$ $5~\rm MeV$.

In Fig.~\ref{F:rhMt}, we display, in the $\rho_B$--$T$ phase diagram, the Mott lines, i.e., the Mott density as a function of temperature, for $d$, $t$, $h$ and $\alpha$, for asymmetric nuclear matter with $\delta$ $=$ $0.13$, obtained employing the phase-space excluded volume approach with ${\bf F}^{\rm cut}$ (i).
As a reference, we include in the figure the Mott densities extracted from an experimental analysis of heavy-ion collisions around the Fermi energy~\cite{HagPRL108}. 
The isospin asymmetry $\delta$ $=$ $0.13$ roughly corresponds to the average isospin asymmetry of the reaction systems, i.e., \isotope[40]{Ar} and \isotope[64]{Zn} bombarding \isotope[112,124]{Sn}, analyzed in Ref.~\cite{HagPRL108}.
In particular, there the Mott densities were deduced by fitting light-cluster emission from the (thermally equilibrated) mid-rapidity region with a statistical model.
}

\begin{figure}[ht]
\centering
\includegraphics[width=6.0cm]{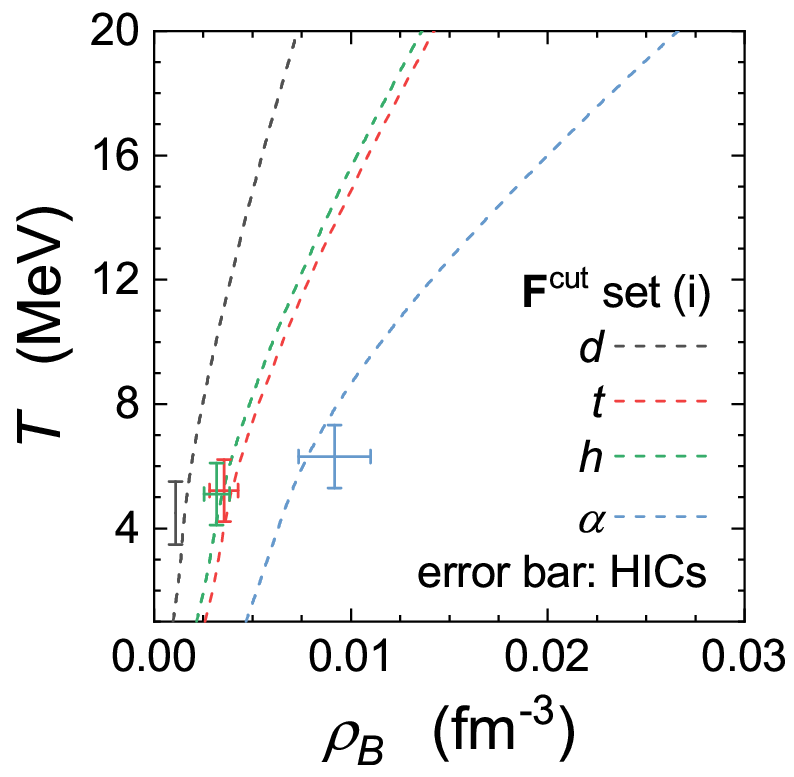}
\caption{Mott lines of $d$, $t$, $h$ and $\alpha$ for nuclear matter with isospin asymmetry $\delta$ $=$ $0.13$ in $\rho_B$--$T$ phase diagram obtained by the phase-space excluded-volume approach with ${\bf F}^{\rm cut}$~(i) $=$ $(0.10,0.15,0.25)$ and the Skyrme interaction MSL$1$.
Error bars represent the Mott densities for light clusters deduced from light-nuclei yields in heavy-ion collisions~(HICs) around the Fermi energy~\cite{HagPRL108}.}
\label{F:rhMt}
\end{figure}

We notice in Fig.~\ref{F:rhMt} that the phase-space excluded-volume approach with ${\bf F}^{\rm cut}$ (i) gives rise to Mott densities of light clusters consistent with those deduced from Fermi-energy heavy-ion collisions~\cite{HagPRL108}.
The latter can be regarded as a probe
of the strength of in-medium effects on light clusters, and can be used to determine a ${\bf F}^{\rm cut}$ suitable for calculations at low baryon densities.
However, they offer limited insight into the strength of the Mott effect and the magnitude of ${\bf F}^{\rm cut}$ at higher densities.
To explore this density range, one may resort to heavy-ion collisions at higher beam energies, where light clusters are primarily produced in warm and dense nuclear matter formed transiently during the compression stage of the reaction~\cite{WRPRC108}.

\subsection{Light-cluster fractions}
\label{S:X}

\begin{figure}[htb]
\centering
\includegraphics[width=\linewidth]{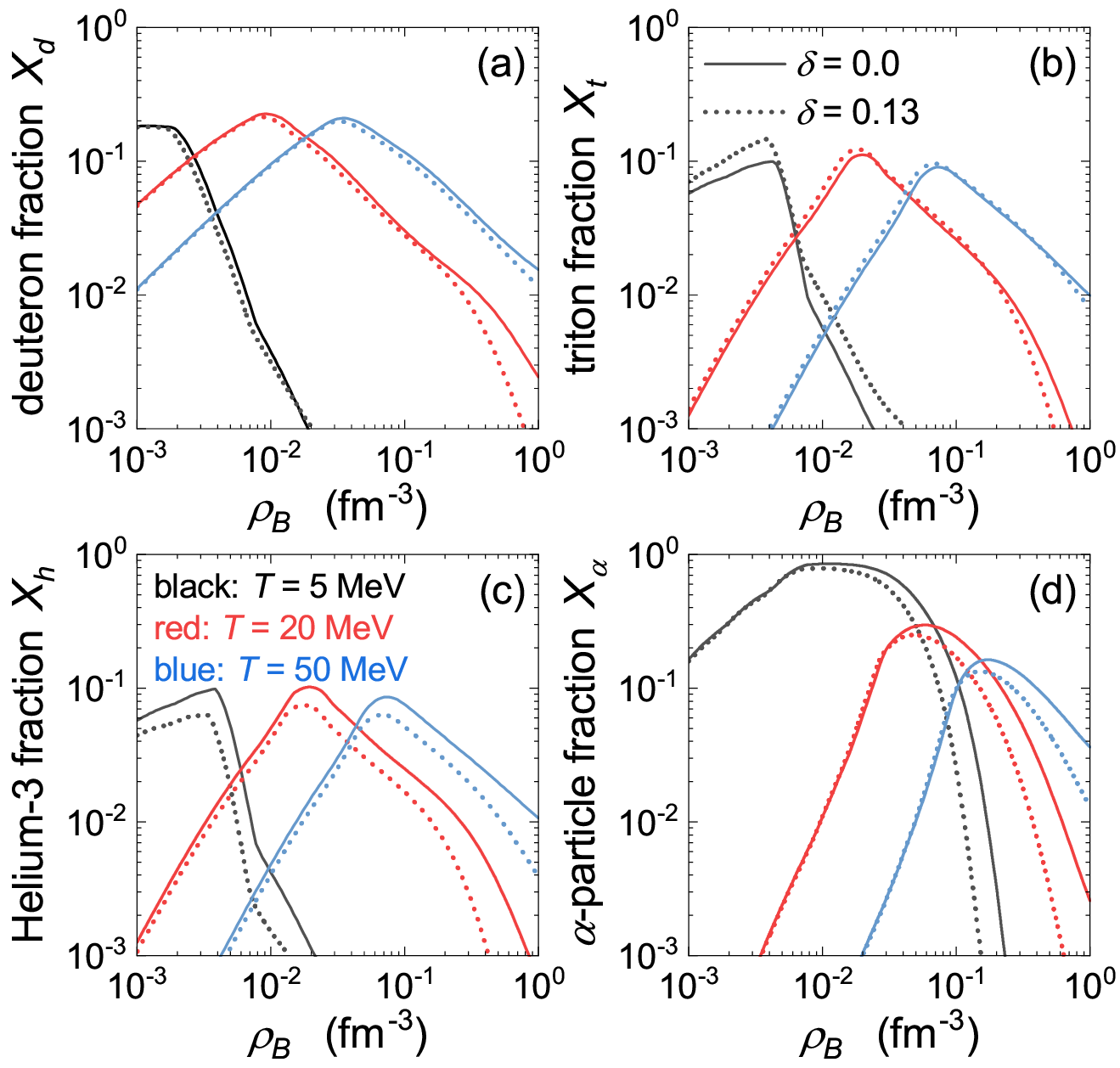}
\caption{Fractions of (a) $d$, (b) $t$, (c) $h$ and (d) $\alpha$ of nuclear matter as functions of baryon density $\rho_B$, calculated using the phase-space excluded-volume approach for two isospin asymmetries at various temperatures.
Results are obtained with the $\bf{F}^{\rm cut}$~(i) $=$ $(0.10,0.15,0.25)$ and the Skyrme interaction MSL$1$.}
\label{F:XLN}
\end{figure}

The light-cluster fractions in nuclear matter $X_{\nu}$ are given by
\begin{equation}
    X_{\nu} = \frac{A}{\rho_B}\int g_{\nu} f_{\nu}(\vec{P})\frac{{\rm d}\vec{P}}{(2\pi\hbar)^3},
\end{equation}
with $g_\nu$ being the spin degeneracy of light-cluster species $\nu$.
We plot in Fig.~\ref{F:XLN} the light-cluster fractions $X_\nu$ of $d$, $t$, $h$ and $\alpha$ for nuclear matter of various densities, at temperature $T$ $=$ $5$, $20$ and $50~\rm MeV$, obtained using the phase-space excluded-volume approach with ${\bf F}^{\rm cut}$~(i) and the Skyrme interaction MSL$1$.
Two isospin asymmetries $\delta$ $=$ $0$ and $0.13$, are considered.

The light-cluster fractions $X_\nu$ presented in Fig.~\ref{F:XLN} exhibit a rise-and-fall pattern.
Below the Mott density, $X_\nu$ increases, in line with the predictions of conventional statistical approaches~\cite{TypPRC81}.
The increase of $X_\nu$ starts to slow down when $\rho_B$ reaches the Mott density, above which the phase space allowed for light clusters to exist is reduced.
Eventually, $X_\nu$ are significantly suppressed in the high-density region.
Another feature emerging from Fig.~\ref{F:XLN} is that the light-cluster fractions in dense nuclear matter increase significantly with increasing temperature.
At high temperature, more nucleons populate high-momentum regions in phase space, where light clusters are allowed to exist.
Therefore, a weaker in-medium suppression of light clusters is observed.
In the case of $\delta$ $=$ $0.13$, the helium-3 fraction is suppressed, as expected, due to a lower $\mu_p$, compared with the symmetric case.
For $d$ and $\alpha$ particles, since their chemical potentials are less affected by the isospin asymmetry, the larger Mott momenta observed in Fig.~\ref{F:PMt} for asymmetric matter lead to the suppression of their fractions, especially for $\alpha$ particles as observed in Fig.~\ref{F:XLN}~(d).

One of the implications of the present work is to propose a connection between the in-medium modifications of light-cluster properties in nuclear matter and the light-nuclei productions in heavy-ion collisions at intermediate energies.
Based on dynamical models, it has been shown that light clusters are mainly produced in the compression stage of the collision~\cite{WRPRC108}.
Therefore, the final-state light-nuclei yields are largely related to the light-cluster fractions in warm and dense nuclear matter, produced transiently in the compression stage.
Therefore, we focus on warm and dense nuclear matter and examine the influence of the strength of the in-medium effect, i.e., the magnitude of $F^{\rm cut}_A$, on light-cluster fractions.

\begin{figure}[ht]
\centering
\includegraphics[width=\linewidth]{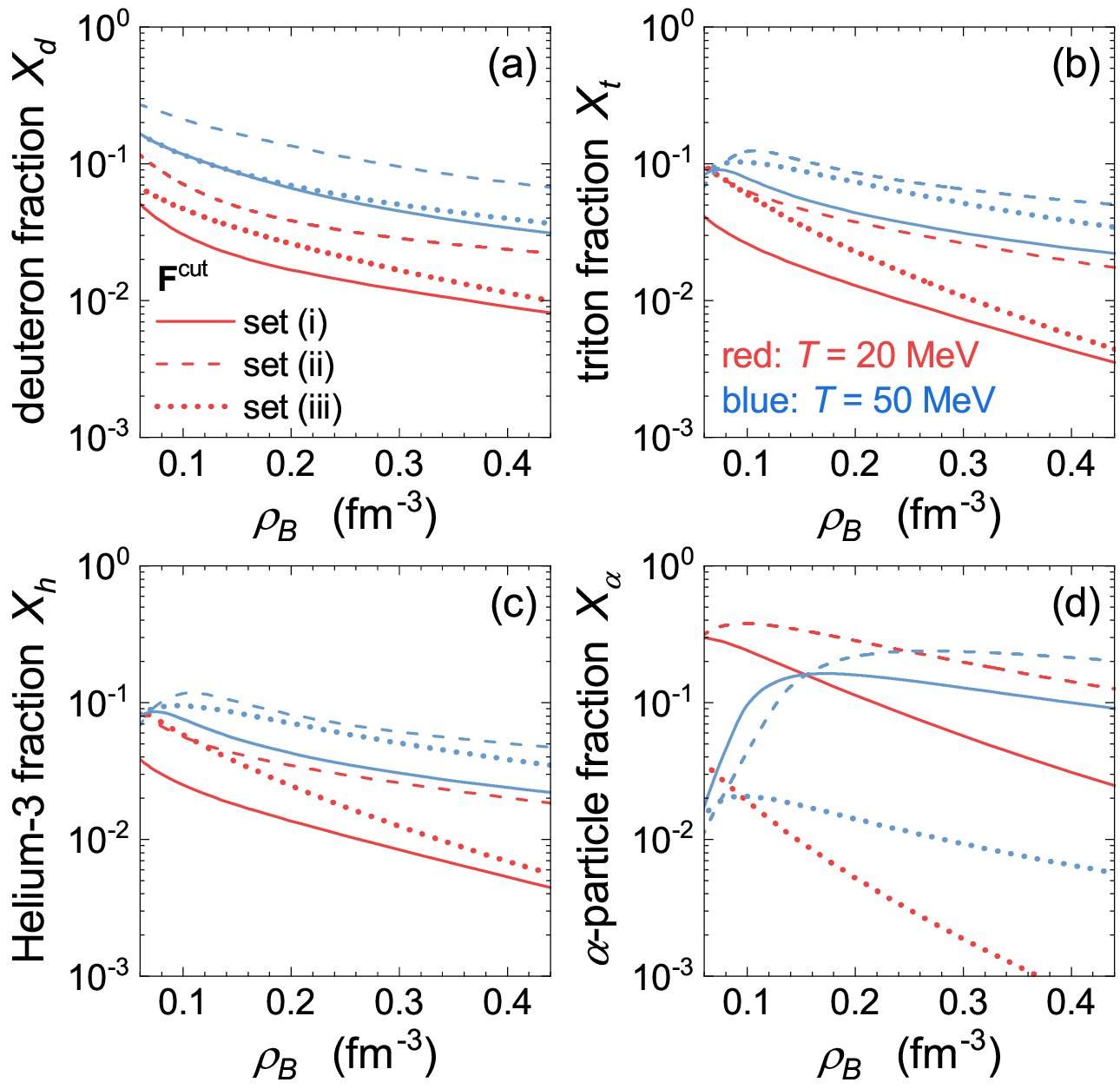}
\caption{Fractions of (a) $d$, (b) $t$, (c) $h$ and (d) $\alpha$ of dense symmetric nuclear matter as functions of baryon density $\rho_B$ at temperature $T$ $=$ $20$ and $50~\rm MeV$ obtained with the phase-space excluded-volume approach.
We employ the Skyrme interaction MSL$1$, and three ${\bf F}^{\rm cut}$ sets in the calculations, namely, the ${\bf F}^{\rm cut}$~(i) $=$ $(0.10,0.15,0.25)$, the ${\bf F}^{\rm cut}$~(ii) $=$ $(0.20,0.25,0.35)$, and the ${\bf F}^{\rm cut}$~(iii) $=$ $(0.10,0.15,0.15)$.}
\label{F:XLN2}
\end{figure}

The fractions of $d$, $t$, $h$ and $\alpha$ in dense symmetric nuclear matter at temperature $T$ $=$ $20~\rm MeV$ and $50~\rm MeV$, obtained with the ${\bf F}^{\rm cut}$~(i), (ii), and (iii), are compared in Fig.~{\ref{F:XLN2}}.
A considerable influence of the magnitude of $F^{\rm cut}_A$ on the light-cluster fractions 
is observed.
The ${\bf F}^{\rm cut}$~(iii), which adopts a smaller $F^{\rm cut}_4$ for $\alpha$ with respect to the ${\bf F}^{\rm cut}$~(i), leads to significantly lower $\alpha$-particle fractions, and subsequently increases slightly the fractions of the other light-cluster species, due to baryon number conservation.
The overall larger ${\bf F}^{\rm cut}$~(ii) results in larger fractions for all light-cluster species at high densities (say, beyond $\rho_0$).
However, in the case of $T$ $=$ $50~\rm MeV$, a suppression of the $\alpha$-particle fraction is observed at low densities, due to the more enhanced deuteron fraction, indicating the interplay between the $F^{\rm cut}_A$ value of different light-cluster species.
This sensitivity to $F^{\rm cut}_A$ has already been observed for the light-nuclei yields in heavy-ion collisions, based on transport models adopting previous formulations of the phase-space excluded-volume approach ~\cite{KuhPRC63,WRPRC108}.
By implementing the present, more refined, approach into dynamical models and comparing with experimental light-nuclei yields in intermediate-energy heavy-ion collisions, the magnitude of $F^{\rm cut}_A$, which represents the strength of the in-medium effects on light clusters, can be deduced , and thus information about the light-cluster fractions in warm and dense nuclear matter can be accessed.
These studies are in progress and will be reported elsewhere.

\subsection{Sensitivity to neutron-proton effective mass splitting}
\label{S:mv}

The concept of nucleon effective mass is widely used to characterize the momentum and/or energy dependence of the single-nucleon potential or the real part of the nucleon self-energy in nuclear medium.
In asymmetric nuclear matter, the difference between neutron and proton potentials may lead to the 
splitting of their effective masses. 
Its dependence on the matter asymmetry is usually characterized by the linear isospin splitting coefficient $\Delta m_{1}^*$ $\equiv$ $\left.\frac{\partial m_{n-p}^*(\rho_0)}{\partial \delta}\right\vert_{\delta =0}$.
Empirically, the isoscalar effective mass $m^*_s$---the nucleon effective mass in symmetric nuclear matter---and $m_{n-p}^*$ can be determined from analyses of finite nuclei structure~\cite{PeaPRC64} and collective motions~\cite{ZZPRC93,KHYPRC95,SJPRC101,ZZCPC45,SYDPRC104}, and nuclear reactions~\cite{CouPRC94,WFYNST34,TsaPLB853,CozPRC110} using mean-field and/or dynamical models, as well as optical model analyses of nucleon–nucleus scattering data~\cite{LXHPLB743}.
One may refer to Ref.~\cite{LBAPPNP99} for a recent review on nucleon effective masses.

Since the effective mass enters the phase-space distribution function, the isoscalar effective mass, as well as the linear isospin splitting coefficient $\Delta m_1^*$ influence the occupations of different particle species.
We concentrate here on the effect of the isospin splitting coefficient, which is currently more debated.
We employ Skyrme interactions with distinct $\Delta m_1^*$ values, namely,  MSL$1$ and MSL$1$-m, to explore this $\Delta m_1^*$-sensitivity of various properties of nuclear matter with light clusters.
At the nuclear saturation density $\Delta m^*_1$ $=$ $0.23m$ and $-0.17m$ for MSL$1$ and MSL$1$-m, respectively.
Details of the nucleon effective masses with the standard Skyrme energy-density functional can be found in Appendix~\ref{SA:Sky}.
In the left window of Fig.~\ref{F:mv}, 
we show the light-cluster fractions $X_\nu$ as functions of baryon density $\rho_B$ for asymmetric nuclear matter with $\delta$ $=$ $0.13$ at temperature $T$ $=$ $20~\rm MeV$, obtained with the phase-space excluded-volume approach employing the Skyrme interaction MSL$1$ and MSL$1$-m.
In these calculations, the $\bf{F}^{\rm cut}$ set (i) is adopted.
It is seen from the figure that the influence of $\Delta m_1^*$ on particle fractions is negligible, due to the fixed isospin asymmetry.
The same situation also applies to the obtained Mott momenta.

\begin{figure}[th]
\centering
\includegraphics[width=\linewidth]{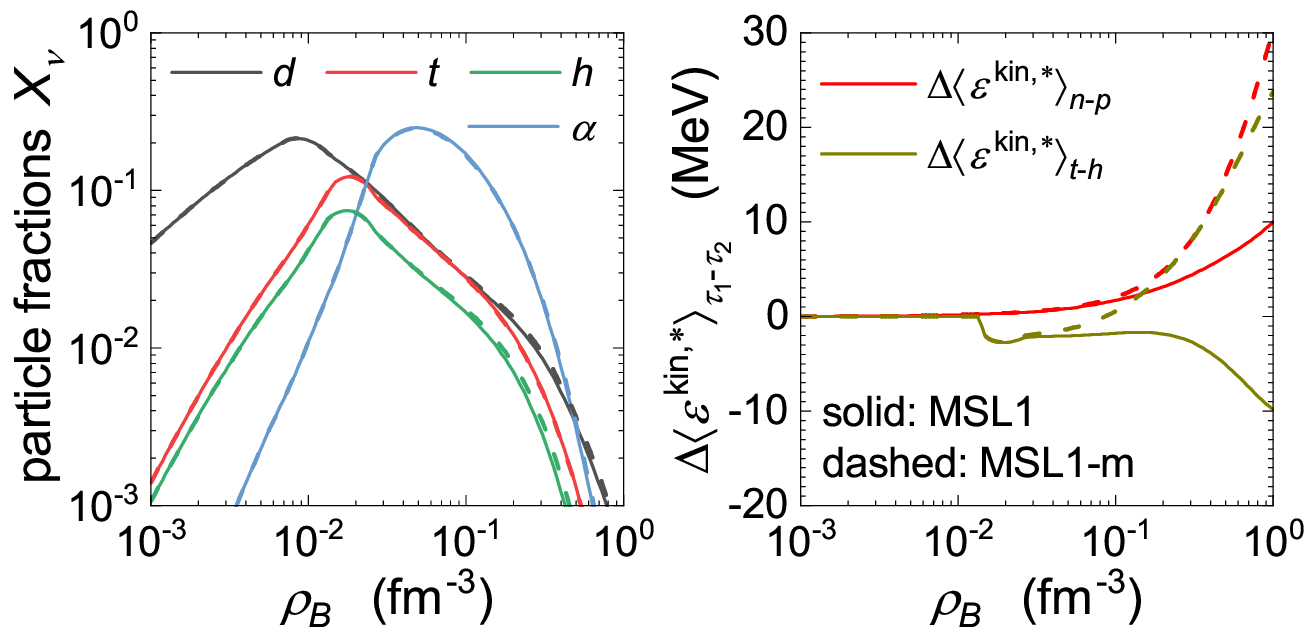}
\caption{The light-cluster fractions $X_\nu$ (left) and the difference between the average single-particle effective kinetic energies of two particle species $\Delta\langle \epsilon^{\rm kin,*}\rangle_{\tau_1-\tau_2}$ (right) as functions of $\rho_B$ for nuclear matter with isospin asymmetry $\delta$ $=$ $0.13$ at temperature $T$ $=$ $20~\rm MeV$, obtained with the phase-space excluded-volume approach.
The $\bf{F}^{\rm cut}$~(i) $=$ $(0.10,0.15,0.25)$, and two Skyrme interactions, i.e., MSL$1$ and MSL$1$-m are employed. 
The non-smooth behavior of $\Delta\langle \epsilon^{\rm kin,*}\rangle_{t-h}$ in the right window is caused by the emergence of non-zero Mott momentum.
}
\label{F:mv}
\end{figure}

We further examine the influence of $\Delta m_1^*$ on the average single-particle effective kinetic energies $\langle \epsilon^{\rm kin,*}\rangle$ of certain particle species.
The difference in $\langle \epsilon^{\rm kin,*}\rangle$ between $n$ and $p$, i.e., $\Delta\langle \epsilon^{\rm kin,*}\rangle_{n-p}$, and that between $t$ and $h$, i.e., $\Delta\langle \epsilon^{\rm kin,*}\rangle_{t-h}$ are presented, as functions of $\rho_B$, in the right window of Fig.~\ref{F:mv}.
The figure exhibits two interesting features.
In the first, $\Delta\langle \epsilon^{\rm kin,*}\rangle_{t-h}$ shows greater sensitivity to $\Delta m_1^*$ than $\Delta\langle \epsilon^{\rm kin,*}\rangle_{n-p}$.
Moreover, a large negative value of $\Delta\langle \epsilon^{\rm kin,*}\rangle_{t-h}$ is observed in high-density region in the case of the MSL$1$.
These two phenomena can be understood as follows.

The average single-particle effective kinetic energies are roughly related to the average single-particle kinetic energies by $\langle \epsilon^{\rm kin,*}\rangle$ $\approx$ $\frac{m}{m^*}\langle \epsilon^{\rm kin}\rangle$ (exact in the non-relativistic limit).
Noting that the dependence of $\langle \epsilon^{\rm kin}\rangle$ on $\Delta m^*_1$ is relatively small,
for a fixed $m^*_s$, changing $\Delta m^*_1$ from negative to positive values leads to a decrease of $\langle \epsilon^{\rm kin,*}\rangle$ for $n$ and $t$, and an increase of $\langle \epsilon^{\rm kin,*}\rangle$ for $p$ and $h$.
Therefore, the positive $\Delta m^*_1$ of MSL$1$ results in  the reduction of both $\Delta\langle \epsilon^{\rm kin,*}\rangle_{n-p}$ and $\Delta\langle \epsilon^{\rm kin,*}\rangle_{t-h}$, with respect to the case of MSL$1$-m (where $\Delta m^*_1$ is negative), 
as can be observed in the right window of Fig.~\ref{F:mv}.
For tritons and helium-3, their $\langle \epsilon^{\rm kin}\rangle$ are enhanced due to their non-zero Mott momenta (see Fig. \ref{F:PMt}).
Accordingly, the deviation of $\Delta\langle \epsilon^{\rm kin,*}\rangle_{t-h}$ between the case MSL$1$ and MSL$1$-m is amplified, compared with that of $\Delta\langle \epsilon^{\rm kin,*}\rangle_{n-p}$.
This amplification is already pronounced at nuclear saturation density, where the triton and helium-3 fractions remain non-negligible.
It therefore indicates that observables of tritons and helium-3 (for instance in heavy-ion collisions) might be more efficient probes for constraining $\Delta m^*_1$ or neutron-proton effective mass splitting than those of neutrons and protons.

The second feature, i.e., the large negative value of $\Delta\langle \epsilon^{\rm kin,*}\rangle_{t-h}$ for MSL$1$ can be understood similarly.
At low densities (say, below $0.1~\rm fm^{-3}$) where the influence of the effective mass is still less pronounced, for both interactions, the magnitude of $\langle\epsilon^{\rm kin,*}\rangle$ for $h$ is larger than that for $t$.
This is in contrast with the behavior observed for $n$ and $p$, which arises from the larger $\mu_n$ with respect to $\mu_p$ in neutron-rich matter (resulting in a larger $\langle\epsilon^{\rm kin}\rangle_n$ compared with $\langle\epsilon^{\rm kin}\rangle_p$).
This observation can be attributed to a larger Mott momentum of $h$ compared with $t$ at low densities, which drives $h$ to populate higher-momentum regions in phase space.
This larger Mott momentum of $h$ (though barely noticeable in Fig.~\ref{F:PMt}) stems from the more compact momentum distribution of the constituent nucleons within $h$ ($\sigma_h$ $>$ $\sigma_t$).
For MSL$1$-m, $\Delta\langle \epsilon^{\rm kin,*}\rangle_{t-h}$ changes its sign at higher densities due to the negative $\Delta m^*_1$.
In contrast, for MSL$1$, because of the positive $\Delta m^*_1$, the $\Delta\langle \epsilon^{\rm kin,*}\rangle_{t-h}$ remains negative.
The boost of the magnitude of $\Delta\langle \epsilon^{\rm kin,*}\rangle_{t-h}$ is again due to the enhanced $\langle\epsilon^{\rm kin}\rangle_\nu$ of $t$ and $h$ caused by the Mott effect (recall the rough relation $\langle \epsilon^{\rm kin,*}\rangle$ $\approx$ $\frac{m}{m^*}\langle \epsilon^{\rm kin}\rangle$).
This large negative value of $\Delta\langle \epsilon^{\rm kin,*}\rangle_{t-h}$ obtained with interactions featuring a positive $m^*_{n-p}$ is in line with observations in heavy-ion collisions often referred to as the ``\isotope[3]{He}-puzzle"~\cite{PogNPA586,LisPRL75,WYJNST36}, i.e., the average kinetic energy of the emitted helium-3 is larger than that of tritons.
That is to say, the in-medium effects on tritons and helium-3, along with a positive $m^*_{n-p}$, may offer a potential resolution to this puzzle.
Detailed studies along this line will be pursued in the future. 

\section{Comparison with in-medium Schr\"odinger equation}
\label{S:IMSE}

To compare the phase-space excluded-volume approach with the in-medium Schr\"odinger equation in describing light cluster properties—such as the Mott momentum—we focus on the two-body case, thus restricting our analysis to nuclear matter composed only of nucleons and deuterons.
We then solve Eq.~(\ref{E:IMSE}) and compare the resulting Mott momentum and deuteron fraction with those obtained from the phase-space excluded-volume approach.
When solving the in-medium Schr\"odinger equation, as usual, the uncorrelated-medium assumption is adopted, i.e., 
the total nucleon occupations $f^{\rm tot}_\tau(\vec{p})$ in Eq.~(\ref{E:IMSE}) are approximated by a Fermi-Dirac distribution,
\begin{equation}
    f^{\rm tot}_\tau(\vec{p}) \approx \tilde{f}^{\rm tot}_\tau(\vec{p}) \equiv \frac{1}{{\rm exp}\big\{\beta[\epsilon_{\tau}(\vec{p})-\tilde{\mu}_{\tau}]\big\} + 1},
\label{E:fFG}
\end{equation}
with $\tilde{\mu}_{\tau}$ determined by requiring that $\int g\tilde{f}^{\rm tot}_\tau(\vec{p})\frac{{\rm d}\vec{p}}{(2\pi\hbar)^3}$ 
yields the {\it total} nucleon density $\rho^{\rm tot}_\tau$.
For the single-nucleon quasiparticle energies in Eq.~(\ref{E:IMSE}), in this section, we approximate the effective mass by the bare nucleon mass, for simplicity.
As for the two-body potential in Eq.~(\ref{E:IMSE}), we choose a simple Gaussian-form separable potential,
\begin{equation}
\begin{split}
    V_{np}(\vec{p}_i,&\vec{p}_j;\vec{p}_i^{~\prime},\vec{p}_j^{~\prime}) = \lambda_d\delta(\vec{p}_i^{~\prime}+\vec{p}_j^{~\prime}-\vec{p}_i-\vec{p}_j)\\
    &\times {\rm exp}\Big[-\frac{(\vec{p}_i-\vec{p}_j)^2}{4\gamma_d^2\hbar^2}\Big]{\rm exp}\Big[-\frac{(\vec{p}_i^{~\prime}-\vec{p}_j^{~\prime})^2}{4\gamma_d^2\hbar^2}\Big],
\end{split}
\end{equation}
with $\lambda_d$ $=$ $-1.2874~\rm GeV\cdot fm^3$ and $\gamma_d$ $=$ $1.474~\rm fm^{-1}$ determined by the binding energy and the r.m.s radius of deuterons in free space~\cite{RopNPA867}.

By means of Eq.~(\ref{E:fFG}), the in-medium Schr\"odinger equation Eq.~(\ref{E:IMSE}) is decoupled from Eqs.~(\ref{E:ftot})--(\ref{E:fpclu}), and becomes manageable to solve to obtain the Mott momentum $P^{\rm Mott}_d$.
Similarly, within the phase-space excluded-volume approach, $P^{\rm Mott}_d$ can be obtained solely from Eq.~(\ref{E:Fcut}).
For given $P^{\rm Mott}_d$, one then determines the chemical potentials, imposing that, from $f_\tau(\vec{p})$ in Eq.~(\ref{E:fN}) and $f_d(\vec{P})$ in Eq.~(\ref{E:fclu}), the total nucleon density $\rho^{\rm tot}_\tau$ is obtained, under the chemical equilibrium condition. Subsequently the particle fractions are evaluated.
In this process, we further neglect the $f^{\rm clu}_\tau(\vec{p})$ in Eq.~(\ref{E:fN}) for simplicity.
We will discuss the validity of this approximation, as well as the uncorrelated-medium assumption, in Appendix~\ref{SA:approx}.
Note that the above two approximations are only adopted in the calculations presented in this Section.

\begin{figure}[htb]
\centering
\includegraphics[width=\linewidth]{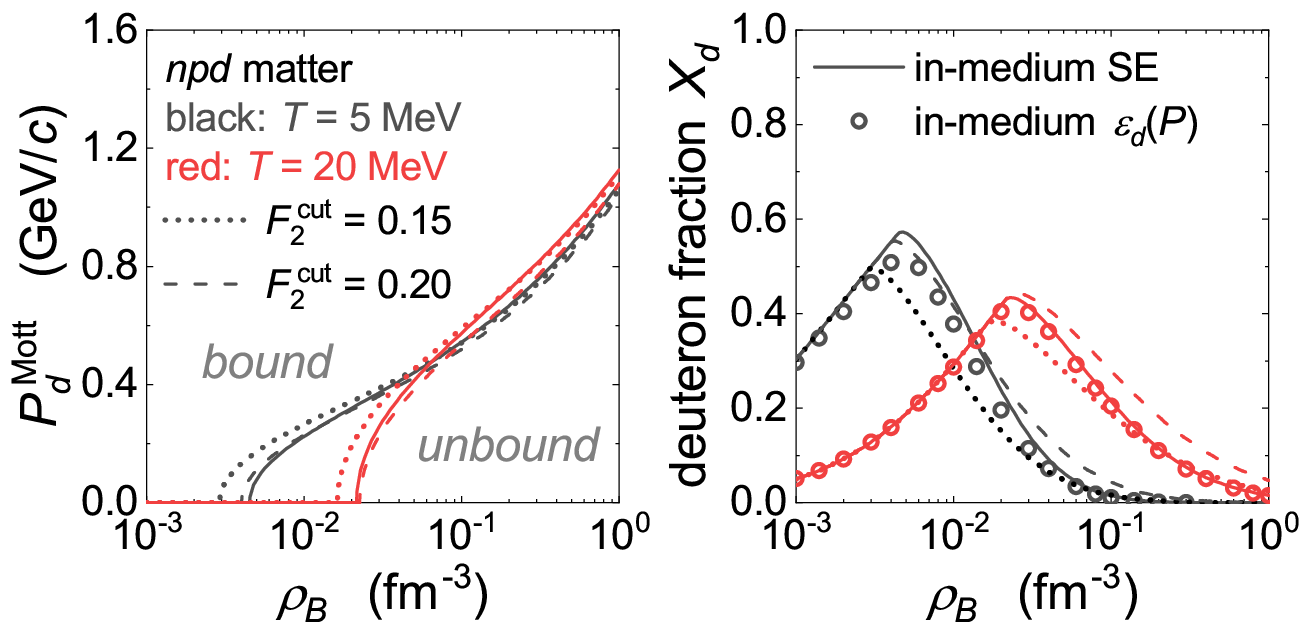}
\caption{The deuteron Mott momentum $P^{\rm Mott}_d$ (left) and fraction $X_d$ (right) as functions of $\rho_B$ for symmetric $npd$ matter at two different temperatures, obtained by solving the in-medium Schr\"odinger equation [in-medium SE and in-medium $\epsilon_d(P)$], and from the phase-space excluded-volume approach with two different $F^{\rm cut}_2$.}
\label{F:cf.IMSE}
\end{figure}

We present in Fig.~\ref{F:cf.IMSE} the deuteron Mott momentum $P^{\rm Mott}_d$ and fraction $X_d$, obtained within the two approaches, as functions of $\rho_B$, for nuclear matter at temperature $T$ $=$ $5~\rm MeV$ and $20~\rm MeV$. 
In the case of the phase-space excluded-volume approach, two $F^{\rm cut}_2$ values are employed.
For the in-medium Schr\"odinger equation, we consider two cases, the one (circles) that includes the in-medium correction on the single particle energy $\epsilon_d(\vec{P})$
entering Eq.(14), and the other one (solid lines) where we simply employ
the free-space binding energy, 
as done in the phase-space excluded-volume approach.
The overall effect of the in-medium correction of $\epsilon_d(\vec{P})$ is not that pronounced, especially in the high-temperature case.
Comparing the results from the two approaches, it is found that, for both temperatures, the obtained $P^{\rm Mott}_d$ and $X_d$ from the in-medium Schr\"odinger equation tend to be better reproduced by a larger $F^{\rm cut}_2$ for small $\rho_B$, 
and a smaller $F^{\rm cut}_2$ for larger $\rho_B$.
Since this density dependence of $F^{\rm cut}_2$ is not that remarkable, a rough overall agreement in terms of $P^{\rm Mott}_d$ and $X_d$ can still be achieved for both temperatures with a constant value of $F^{\rm cut}_2$ lying in between $0.15$ and $0.20$.
In addition, in certain nuclear processes, e.g., heavy-ion collisions, one typically does not need to consider nuclear matter over such a wide density range spanning several orders of magnitude, which also legitimizes the use of a constant $F^{\rm cut}_A$.

In the end, we want to mention that the results obtained from solving the in-medium Schr\"odinger equation, displayed in Fig.~\ref{F:cf.IMSE}, should not be regarded as a strict reference, since they actually depend on the choice of the many-body interaction $V(1,2,\dots,A;1',2',\cdots,A')$.
For a realistic, non-separable potential, the equation becomes increasingly complicated in nuclear matter, let alone in dynamical processes.
In this sense, the empirical information deduced from experiments, based on effective approaches such as the one presented in this work, is still needed to get access to the in-medium properties of light clusters.

\section{Summary and outlook}
\label{S:smy}
We have developed a phase-space excluded-volume approach for light clusters to study their in-medium properties in nuclear matter.
In this approach, the total nucleon phase-space occupation around a light cluster---with contributions from both unbound nucleons and light clusters---is compared with a cutoff parameter $F^{\rm cut}_A$ to determine whether the light cluster can survive in the nuclear medium.
This criterion is inspired by the more sophisticated in-medium Schr\"odinger equation, but its implementation requires significantly less computational effort.
The parameter $F^{\rm cut}_A$ can be regarded as a surrogate of the strength of in-medium effects on light clusters.
Standard Skyrme energy-density functionals have been used to model the nuclear mean-field potential.
The approach has been employed to calculate the Mott momenta and fractions of light clusters in nuclear matter, accounting for the interplay between in-medium effects and thermodynamic properties.
Our study allows to assess the sensitivity of these in-medium properties to various conditions of  density, temperature, isospin asymmetry, and strength of in-medium effects represented by different $F^{\rm cut}_A$ values.
We have further shown that these quantities agree with those obtained from the in-medium Schr\"odinger equation in a simplified case.
When exploring the kinetic properties of light clusters, it is also found that, in nuclear matter, the difference in the average effective kinetic energies between tritons and helium-3, i.e., $\Delta\langle\epsilon^{\rm kin,*}\rangle_{t-h}$ is quite sensitive to the neutron-proton effective mass splitting $m^*_{n-p}$, 
mainly due to the Mott effect.
A weaker sensitivity is seen for neutron-proton average effective kinetic energies difference $\Delta\langle\epsilon^{\rm kin,*}\rangle_{n-p}$, implying that light-nuclei observables (mainly available in heavy-ion collisions) could be more efficient probes of isospin-dependent quantities than nucleon observables.

As a further development, the present phase-space excluded-volume approach can be employed to calculate the equation of state of nuclear matter with light clusters at both zero and finite temperatures.
Moreover, since the approach can be applied to nuclear medium with arbitrary phase-space distributions, it bridges the in-medium properties of light clusters in nuclear matter, e.g., light-cluster fractions, with those in non-equilibrium processes like heavy-ion collisions.
The present work thus paves the way for probing the in-medium properties of light clusters in warm and dense nuclear matter through intermediate-energy heavy-ion collisions.
These studies will be pursued in the future.

\begin{acknowledgments}
The authors thank Bao-Jun Cai and Stefan Typel for useful discussions.
Zhen Zhang acknowledges the support by the National Natural Science Foundation of China under Grant No. $12235010$.
Rui Wang and Zhen Zhang acknowledge the hospitality of the Shanghai Research Center for Theoretical Nuclear Physics, where part of this work was discussed, as well as the corresponding support by the National Natural Science Foundation of China under Grant No.$12147101$.

\end{acknowledgments}

\appendix
\renewcommand\thefigure{\Alph{section}\arabic{figure}}
\setcounter{figure}{0}
\renewcommand\thetable{\Alph{section}\arabic{table}}
\setcounter{table}{0}

\section{Skyrme energy-density functionals and nucleon effective masses}
\label{SA:Sky}

Within the standard Skyrme energy-density functional~\cite{ChaNPA627,ChaNPA635}, the nucleon single-particle potential $\epsilon^{\rm pot}_\tau(\vec{p})$ can be expressed as
\begin{widetext}
\begin{equation}
\epsilon^{\rm pot}_{\tau}(\vec{p}) = \epsilon^{{\rm pot},0}_\tau(\rho_n,\rho_p) + \frac{C^{[2]}}{8\hbar^2}\int \frac{{\rm d}\vec{p}\,'}{(2\pi\hbar)^3}(\vec{p}-\vec{p}\,')^2g\big[f^{\rm tot}_n(\vec{p}\,')+f^{\rm tot}_p(\vec{p}\,')\big] + \frac{D^{[2]}}{8\hbar^2}\int \frac{{\rm d}\vec{p}\,'}{(2\pi\hbar)^3}(\vec{p}-\vec{p}\,')^2gf^{\rm tot}_{\tau}(\vec{p}\,'),
\label{E:eSky}
\end{equation}
\end{widetext}
with $\epsilon^{{\rm pot},0}_\tau(\rho_n,\rho_p)$ being the density-dependent part, and $C^{[2]}$ and $D^{[2]}$ being combinations of the
Skyrme parameters $t_1$, $x_1$, $t_2$ and $x_2$, defined as
\begin{eqnarray}
    C^{[2]} & = & t_1(x_1+2)+t_2(x_2+2)\label{E:C2},\\
    D^{[2]} & = & -t_1(2x_1+1)+t_2(2x_2+1)\label{E:D2}.
\end{eqnarray}
Note that $\rho_{\tau}$ in Eq.~(\ref{E:eSky}) should be understood as $\rho^{\rm tot}_{\tau}$.
The advantage of employing the standard Skyrme interaction is that, in nuclear matter, the momentum dependence of $\epsilon^{\rm pot}_\tau$ can be recast into the kinetic part $\epsilon^{\rm kin}_\tau$ by replacing the mass with the effective mass $m^*$.
This can be seen by expanding the integrand in $C^{[2]}$ and $D^{[2]}$ terms, 
\begin{equation}
    \int \frac{{\rm d}\vec{p}\,'}{(2\pi\hbar)^3}(\vec{p}-\vec{p}\,')^2gf^{\rm tot}_{\tau}(\vec{p}\,') = p^2\rho^{\rm tot}_{\tau} + \langle p^2\rangle_{\tau}.
\label{E:MDNM}
\end{equation}
Note that the $\vec{p}\cdot\vec{p}\,'$ term vanishes because of the spherical symmetry of the momentum distribution in nuclear matter.
Since the $p$-dependent part, i.e.,  $p^2\rho^{\rm tot}_{\tau}$, does not explicitly depend on $f^{\rm tot}_\tau(\vec{p})$, it can be combined with the kinetic part $\epsilon^{\rm kin}_\tau$ to define the effective kinetic energy $\epsilon^{\rm kin,*}_\tau$.
With Eq.~(\ref{E:eSky}), the nucleon effective mass in Eq.~(\ref{E:Em}) can be expressed as
\begin{equation}
\begin{split}
    \frac{m}{m^*_\tau} &\equiv 1 + m\frac{\vec{p}\cdot\nabla_{\vec{p}}\epsilon^{\rm pot}_\tau(\vec{p})}{p^2}\\
    & = 1 + \frac{m}{4\hbar^2}\big[C^{[2]}(\rho^{\rm tot}_n+\rho^{\rm tot}_p) + D^{[2]}\rho^{\rm tot}_{\tau}\big].
\end{split}
\label{E:Em2}
\end{equation}
Effective masses of light clusters expressed in Eq.~(\ref{E:EmLN}) can be obtained similarly.
It can be seen that, for standard Skyrme interactions, the effective mass only depends on the (total) nucleon density and not on the single-particle momentum. 
More generally, for Skyrme interactions extended to include higher-order derivative terms (relative-momentum terms)~\cite{CarPRC78,RaiPRC83,WRPRC98,WSPPRC109,YJPRC109,WSPPRC111,YJTApJ985}, this condition does not hold any more.

The Skyrme parameters of MSL$1$ and MSL$1$-m used in the present work are provided in Table~\ref{T:MSL}.
Note that only parameters $C^{[2]}$ and $D^{[2]}$ given in Eqs.~(\ref{E:C2}) and (\ref{E:D2}), respectively, are relevant 
for the calculations presented in this work.
For the corresponding macroscopic quantities~\cite{CLWPRC82}, such as the incompressibility coefficient $K_0$ and the symmetry energy $E_{\rm sym}(\rho_0)$ at nuclear saturation density, one may refer to Ref.~\cite{ZZPLB726}.
These quantities given by MSL$1$ and MSL$1$-m differ only in the isoscalar effective mass $m^*_{v,0}$ (related to $\Delta m^*_1$).

\begin{table}[t]
\caption{Skyrme parameters of MSL$1$ and MSL$1$-m. Note that only two free parameters among $t_1$, $x_1$, $t_2$ and $x_2$ are needed for describing nuclear matter, namely, their combinations $C^{[2]}$ and $D^{[2]}$ given in Eqs.~(\ref{E:C2}) and (\ref{E:D2}), respectively.}
\begin{tabular}{ccc}
    \hline\hline
    Skyrme parameter & ~~~~~~MSL$1$~~~~~~ & ~~~~~~MSL$1$-m~~~~~~ \\
    \hline
    $t_0$ $({\rm MeV\cdot fm^3})$ & $-1963.23$ & $-1962.93$ \\
    $t_1$ $({\rm MeV\cdot fm^5})$ & $379.845$ & $379.795$\\
    $t_2$ $({\rm MeV\cdot fm^5})$ & $-394.554$ & $-714.113$\\
    $t_3$$({\rm MeV\cdot fm}^{3+3\sigma})$ & $12174.9$ & $12174.3$\\
    $x_0$ & $0.320770$ & $0.529806$ \\
    $x_1$ & $0.344849$ & $0.134762$ \\
    $x_2$ & $-0.847304$ & $-1.02737$ \\
    $x_3$ & $0.321930$ & $0.865856$ \\
    $\sigma$ & $0.269359$ & $0.269417$ \\
    \hline\hline
\end{tabular}
\label{T:MSL}
\end{table}

\section{Approximations on total nucleon occupation}
\label{SA:approx}

As we have mentioned in the main text, an uncorrelated-medium assumption, i.e., replacing the total nucleon occupation $f^{\rm tot}_\tau(\vec{p})$ in the in-medium Schr\"odinger equation Eq.~(\ref{E:IMSE}) with a Fermi-Dirac distribution $\tilde{f}^{\rm tot}_\tau(\vec{p})$ given in Eq.~(\ref{E:fFG}), is usually adopted to reduce computational complexity.
With the uncorrelated-medium assumption, one can first obtain the Mott momenta of light clusters solely from Eq.~(\ref{E:IMSE}) or Eq.~(\ref{E:PMt}), and then treat the nuclear matter as normal thermal-equilibrium problems with different particle species.

\begin{figure}[b]
\centering
\includegraphics[width=\linewidth]{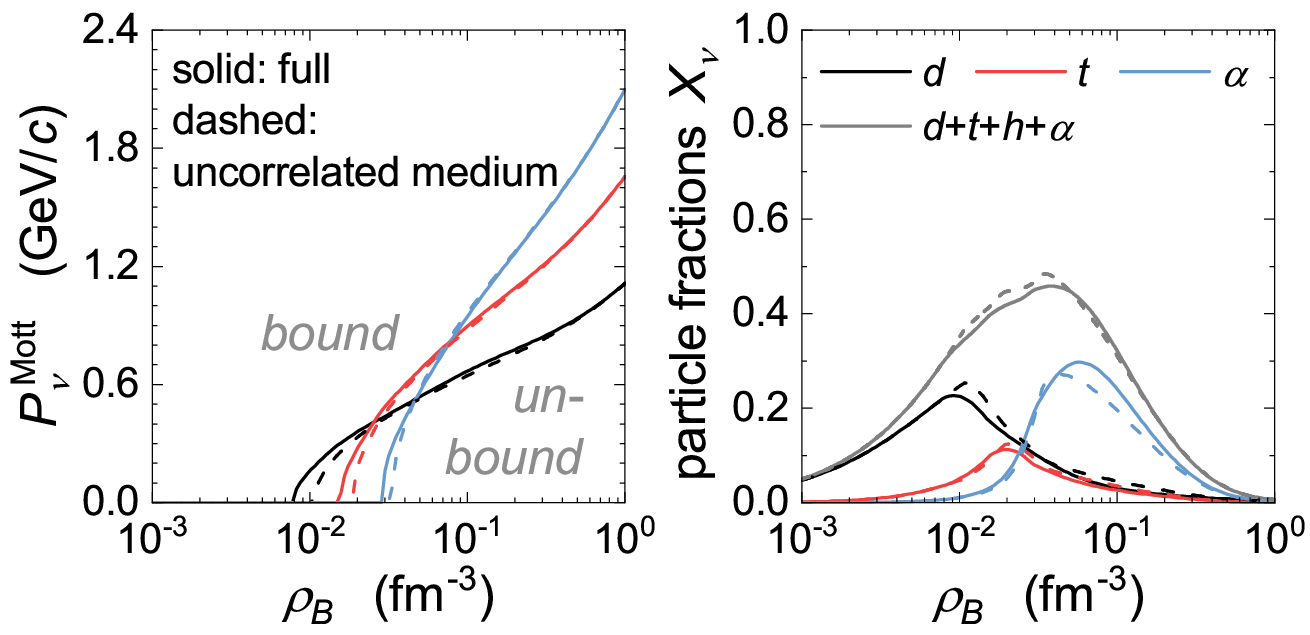}
\caption{Mott momenta $P^{\rm Mott}_\nu$ (left) and fractions $X_\nu$ (right) of light clusters of symmetric nuclear matter as functions of baryon density $\rho_B$ at temperature $T$ $=$ $20~\rm MeV$ obtained using the phase-space excluded-volume approach with ${\bf F}^{\rm cut}$~(i) $=$ $(0.10,0.15,0.25)$ and the Skyrme interaction MSL$1$.
Dashed lines represent the results where the uncorrelated-medium assumption is employed.
}
\label{F:UCmd}
\end{figure}

In order to examine the validity of this assumption, we employ it in the phase-space excluded-volume approach, i.e., replacing the $f^{\rm tot}_\tau(\vec{p})$ in Eq.~(\ref{E:Fcut}) with $\tilde{f}^{\rm tot}_\tau(\vec{p})$, and compare the obtained Mott momenta $P^{\rm Mott}_\nu$ and fractions $X_\nu$ of light clusters with those from the full calculations presented in Secs.~\ref{S:PMt} and \ref{S:X}.
The comparison is shown in Fig.~\ref{F:UCmd} for $d$, $t$ and $\alpha$.
Calculations are performed for symmetric nuclear matter at temperature $T$ $=$ $20~\rm MeV$, with ${\bf F}^{\rm cut}$~(i) and the Skyrme interaction MSL$1$.
The results for the full phase-space excluded-volume calculation are the same as those in Figs.~\ref{F:PMt} and \ref{F:XLN}.
It is seen from the figure that the application of the uncorrelated-medium assumption only causes slight deviations with respect to the full calculations for both $P^{\rm Mott}_\nu$ and $X_\nu$.


\begin{figure}[h]
\centering
\includegraphics[width=\linewidth]{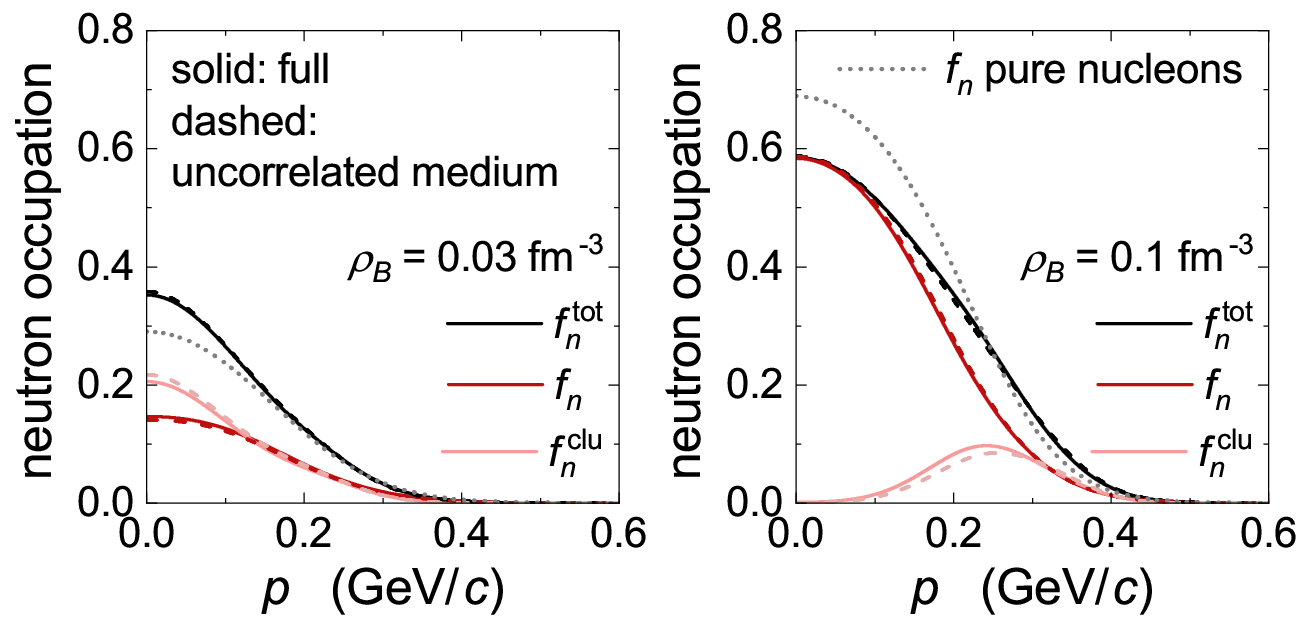}
\caption{Corresponding neutron occupations $f^{\rm tot}_n(\vec{p})$, $f_n(\vec{p})$ and $f^{\rm clu}_n(\vec{p})$ for the calculations shown in Fig.~\ref{F:UCmd} for two different $\rho_B$.
The gray dotted lines represent $\tilde{f}^{\rm tot}_\tau(\vec{p})$, or the neutron distribution in pure nucleon matter at the indicated temperature and density.}
\label{F:UCmdf}
\end{figure}

We then go into the detailed nucleon occupations in momentum space obtained within the phase-space excluded-volume approach.
We present in Fig.~\ref{F:UCmdf} the total neutron occupation $f^{\rm tot}_n(\vec{p})$ [Eq.~(\ref{E:ftotNM})], that from unbound neutrons $f_n(\vec{p})$ [Eq.~(\ref{E:fN})], and that from neutrons bound in light clusters $f^{\rm clu}_n(\vec{p})$ [Eq.~(\ref{E:fnclu})], corresponding to the calculations shown in Fig.~\ref{F:UCmd} at two baryon densities.
Also shown in the figure is the Fermi-Dirac distribution $\tilde{f}^{\rm tot}_\tau(\vec{p})$ in Eq.~(\ref{E:fFG}) employed in the uncorrelated-medium assumption (or the nucleon distribution in pure nucleon matter), whose difference from the exact $f^{\rm tot}_\tau(\vec{p})$ is clearly seen.
Since $\tilde{f}^{\rm tot}_\tau(\vec{p})$ roughly coincides with the exact $f^{\rm tot}_\tau(\vec{p})$ when they become small, to which the Mott momenta are more relevant, they finally lead to similar Mott momenta and fractions of light clusters in nuclear matter, as seen in Fig.~\ref{F:UCmd}.
This finding justifies to some extent approximating $f^{\rm tot}_\tau(\vec{p})$ by $\tilde{f}^{\rm tot}_\tau(\vec{p})$ in the in-medium Schr\"odinger equation.
Nonetheless, the uncorrelated-medium assumption is applicable only in calculations of nuclear matter.
In non-equilibrium processes, $f^{\rm tot}_\tau(\vec{p})$ should be evaluated self-consistently, most likely numerically, depending on specific conditions.
It is worth mentioning that, although the deviation between $\tilde{f}^{\rm tot}_\tau(\vec{p})$ and the exact $f^{\rm tot}_\tau(\vec{p})$ results in similar light-cluster fractions, it can cause a sizable difference in average kinetic energy of nuclear matter, indicating the necessity of including nucleon clustering when evaluating the nuclear equation of state.

The other point we want to discuss concerns the correction on the occupation of unbound nucleons from light clusters.
In Eq.~(\ref{E:fN}), we propose that the occupation of unbound nucleons in nuclear matter should be modified due to the presence of light clusters and thus deviate from the Fermi-Dirac distribution, since in principle the Pauli blocking effect on unbound nucleons from constituent nucleons in light clusters should also be taken into account.
To determine the nucleon chemical potential $\mu_\tau$ by normalizing Eq.~(\ref{E:ftotNM}) usually involves iteration procedures.
The presence of the additional term $f^{\rm clu}_\tau(\vec{p})$ in the nominator of Eq.~(\ref{E:fN}) obliges one to calculate the integrals in Eqs.~(\ref{E:fnclu}) and (\ref{E:fpclu}) for each trial $\mu_\tau$, causing substantial extra computational effort.
We therefore examine whether this additional term plays an important role in calculating light-cluster fractions.

\begin{figure}[t]
\centering
\includegraphics[width=\linewidth]{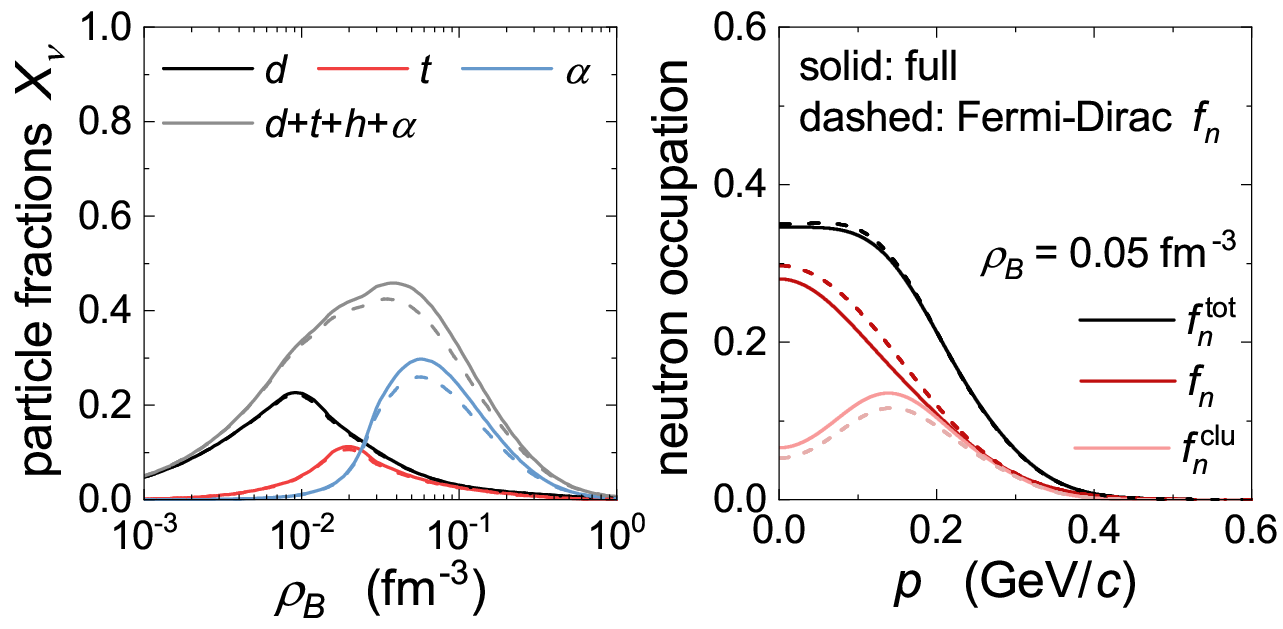}
\caption{Left: light-cluster fractions $X_\nu$ of symmetric nuclear matter as functions of baryon density $\rho_B$ at temperature $T$ $=$ $20~\rm MeV$ using the phase-space excluded-volume approach with the ${\bf F}^{\rm cut}$~(i) $=$ $(0.10,0.15,0.25)$ and the Skyrme interaction MSL$1$.
Right: the corresponding neutron occupations $f^{\rm tot}_n(\vec{p})$, $f_n(\vec{p})$ and $f^{\rm clu}_n(\vec{p})$ at $\rho_B$ $=$ $0.05~\rm fm^{-3}$.
Dashed lines represent the results where the $f^{\rm clu}_\tau(\vec{p})$ in Eq.~(\ref{E:fN}) is neglected.
}
\label{F:fNFD}
\end{figure}

We calculate the light-cluster fractions in nuclear matter at $T$ $=$ $20~\rm MeV$ using the ${\bf F}^{\rm cut}$~(i) and the Skryme interaction MSL$1$ (same as those in Fig.~\ref{F:UCmd}), employing the phase-space excluded-volume approach but neglecting the $f^{\rm clu}_\tau(\vec{p})$ in Eq.~(\ref{E:fN}) [imposing a Fermi-Dirac $f_\tau(\vec{p})$].
Note the distinction between this Fermi-Dirac $f_\tau(\vec{p})$ from $\tilde{f}^{\rm tot}_\tau(\vec{p})$, with the former representing the occupation of unbound nucleons and the latter being an approximation of the total nucleon occupation.
Apart from the light-cluster fractions $X_\nu$ given in the left window of Fig.~\ref{F:fNFD}, we show in the right window the corresponding neutron occupations $f^{\rm tot}_n(\vec{p})$, $f_n(\vec{p})$ and $f^{\rm clu}_n(\vec{p})$ at $\rho_B$ $=$ $0.05~\rm fm^{-3}$.
The correction $f^{\rm clu}_\tau(\vec{p})$ in Eq.~(\ref{E:fN}) decreases the magnitude of $f_\tau(\vec{p})$.
It therefore results in a systematic increase in light-cluster fractions, as can be seen in the left window of Fig.~\ref{F:fNFD}. 
On the other hand, the influence of the correction on the total nucleon occupation $f^{\rm tot}_\tau$ is negligible, mainly due to the particle-number conservation.
It therefore means that the Mott momenta obtained with and without the correction are almost identical to each other, which is the reason why we only show the light-cluster fractions in Fig.~\ref{F:fNFD}.

In general, the effect of omitting $f^{\rm clu}_\tau(\vec{p})$ in Eq.~(\ref{E:fN}), as well as adopting the uncorrelated-medium assumption, is not that pronounced in the cases shown here.
However, we would like to mention that these two approximations may break down when the light-cluster fractions are large enough, for example, when larger values of $F^{\rm cut}_A$ are used.



\bibliography{NM_fcut}

\end{document}